\documentclass[aps, physrev, reprint, superscriptaddress,
amsmath, amssymb, floatfix]{revtex4-2}

\usepackage{physics}
\usepackage{graphicx}
\usepackage{float}
\usepackage{accents}
\usepackage[pdftex]{hyperref}
\usepackage{multirow}
\hypersetup{colorlinks=true,linkcolor = blue, anchorcolor = red, citecolor = blue,filecolor = red, urlcolor = blue}
\usepackage[english]{babel}
\newcommand{\e}{\text{e}}
\newcommand{\aop}{\hat{a}}
\newcommand{\aopd}{\hat{a}^\dag}
\newcommand{\bop}{\hat{b}}
\newcommand{\bopd}{\hat{b}^\dag}
\newcommand{\D}{\hat{\mathcal{D}}}
\newcommand{\Ham}{\hat{H}}
\newcommand{\Vop}{\hat{V}}

\newcommand{\sigm}{\hat{\sigma}_-}
\newcommand{\sigp}{\hat{\sigma}_+}

\begin{document}

\title{Bimodal non-Gaussian photonic states from a single quantum emitter in a waveguide}

\newcommand{\AffLKB}{Laboratoire Kastler Brossel, Sorbonne Université, CNRS, ENS-PSL Research University, Collège de France, Paris, France}
\newcommand{\AffWeizmann}{Department of Physics of Complex Systems, Weizmann Institute of Science, Rehovot, 76100, Israel}
\newcommand{\AffPalackyUni}{Department of Optics, Faculty of Science, Palacký University, 771 46 Olomouc, Czech Republic}

\author{Thomas Copie}
\email[Contact author: ]{thomas.copie@lkb.upmc.fr}
\affiliation{\AffLKB}

\author{Ofer Firstenberg}
\affiliation{\AffWeizmann}

\author{G. P. Teja}
\affiliation{\AffPalackyUni}

\author{Radim Filip}
\affiliation{\AffPalackyUni}

\author{Hanna Le Jeannic}
\email[Contact author: ]{hanna.le-jeannic@cnrs.fr}
\affiliation{\AffLKB}

\date{\today}

\begin{abstract}
We investigate the generation of deterministic and heralded non-Gaussian states of light, using a single two-level system coupled to a chiral waveguide. We study the case of a single two level system  driven by pulsed coherent and squeezed drive in a chiral waveguide. For coherent input pulses, we show that the emitter can deterministically generate  Wigner-negative states, albeit of limited rank. Going beyond, using squeezed-vacuum inputs, we show that the interaction produces rich bimodal non-Gaussian states from which higher-stellar-rank states, including large squeezed cat states, can be experimentally extracted with a substantial success rate. For experimental feasibility and comparison, we analyze the effect of imperfect coupling and of the intrinsic $50\%$ collection limit of symmetric, non-chiral waveguides. Following that, we propose a simple interferometric scheme that recovers an effectively chiral interaction in an otherwise bidirectional waveguide. 
\end{abstract}

\maketitle

\section{Introduction}
Photonic continuous-variable quantum computing has made remarkable progress in recent years, with the generation of large entangled states and the demonstration of increasingly complex quantum optical processing protocols \cite{larsen_deterministic_2019,larsen_GKP_2025}. These achievements highlight the scalability and versatility of Gaussian photonic platforms, where propagating squeezed states, linear optics, and homodyne detection can be combined with a high degree of control. At the same time, Gaussian resources alone are not sufficient for universal quantum computation or for the preparation of several key fault-tolerant resource states. Photonic quantum computation requires capabilities that go beyond the Gaussian regime \cite{gottesman1998heisenbergrepresentationquantumcomputers, BartlettGaussianStateSimulation}. A central challenge is therefore to access genuinely non-Gaussian states and operations in a scalable way. This is particularly important for bosonic codes such as Gottesman-Kitaev-Preskill (GKP) states \cite{GKP2001}, which provide a promising route toward fault-tolerant continuous-variable quantum computing. More generally, non-Gaussian states such as photon-number states, photon-number superpositions, and Schrödinger cat states constitute essential resources for optical quantum information processing. 

The distinction between Gaussian and non-Gaussian states is naturally expressed in phase space through the Wigner function. Genuinely quantum non-Gaussian states can exhibit negative regions in their Wigner functions, reflecting interference effects that cannot be captured within a purely Gaussian description. In optical platforms, such states are commonly generated through measurement-induced processes, including photon subtraction, photon addition, or photon-number-resolved heralding. These techniques have enabled important experimental progress, but their probabilistic character poses challenges for scalability and synchronization.

Deterministic generation of non-Gaussian optical states can be pursued by coupling propagating light to a quantum emitter providing an effective few-photon nonlinearity. This idea has first been explored in cavity-QED \cite{haroche_exploring_2006}, where single atoms or atomic ensembles strongly interact with a well-defined optical mode \cite{Honer_2011,Daiss2019}. In such systems, multilevel atomic schemes have enabled the generation of optical cat states conditioned on atomic-state measurements \cite{Hacker2019}, and more recently the deterministic preparation of freely propagating Wigner-negative states encoded in the ($|0\rangle$) and ($|1\rangle$) subspace using an intracavity Rydberg superatom \cite{magro_deterministic_2023}.

With the progress of nanophotonic interfaces, waveguide QED has emerged as a complementary route to few-photon nonlinear optics, replacing a closed cavity mode by highly coupled propagating modes. Several proposals have exploited collective emitter-waveguide interactions to generate single- or multi-mode multiphoton states \cite{Gonzalez-Tudela_2015_Det,Paulisch_2018}. More recent works have explored deterministic photon subtraction and addition in waveguide QED settings, either from Fock state inputs \cite{lund_subtraction_2024}, or by using a $\Lambda$-atom \cite{pasharavesh_generation_2024} or a dynamically coupled two-level system (TLS) \cite{luo2025dynamicstimulatedemissiondeterministic}.
Recent work has also analyzed the temporal-mode structure of light scattered by a two-level system, showing how few-photon inputs can be converted into multimode non-classical output states \cite{Bouchereau2026}.

In parallel, recent studies have shown that one or a few two-level emitters coupled to a waveguide can generate non-Gaussian states of light, either in transient regimes \cite{kleinbeck_creation_2023} or in driven steady-state configurations \cite{quijandria_steady-state_2018, leonhardt_wigner-negative_2025}. More generally, squeezed few-photon superposition states have recently been identified as efficient intermediate resources for optical cat-state generation \cite{Luo_2026}. Building on these developments, we investigate protocols for generating non-Gaussian states of light using a single two-level system coupled to a waveguide as presented in Fig. \ref{fig:virtual_cavities}(a). We focus on a pulsed regime, which complements promising steady-state approaches \cite{quijandria_steady-state_2018,leonhardt_wigner-negative_2025} and consider in particular the regime where only few modes are involved. 

We consider two types of incoming Gaussian states that can be produced deterministically in quantum optics experiments. We first study the case of a coherent-state input pulse, where deterministic non-Gaussian states of light can already be generated. Within the parameter range explored here, the resulting states are found to have a simple non-Gaussian structure, with core states predominantly confined to the $\{\ket{0},\ket{1}\}$ subspace. We then turn to pulsed squeezed-vacuum inputs, for which we find that interesting non-Gaussian resource states such as large squeezed cat states can be produced, either deterministically (albeit impure) or using a quantum pulsed gate, followed by a single photon detection.
In this case, we go beyond photon addition and subtraction \cite{lund_subtraction_2024, pasharavesh_generation_2024, luo2025dynamicstimulatedemissiondeterministic} by making use of the rich multi-modal structure of squeezed light scattered on a TLS. Finally, we analyze how Wigner negativity survives imperfect emitter-waveguide coupling, including the intrinsic 50\% coupling limit of non-chiral configurations. We finally propose a way to always recover chirality in a perfect but non-chiral waveguide-QED system.   

\section{Model}
We consider a two-level system (TLS), with a ground state $\ket{g}$ and an excited state $\ket{e}$ separated by $\hbar \omega_0$, 
coupled to a one-dimensional waveguide. We assume an ideal chiral waveguide, such that light propagates only from left to right \cite{lodahl_chiral_2017}. 
The TLS is thus coupled to a single propagation direction containing a continuum of frequency and temporal modes. 
The free field Hamiltonian is $\Ham_F = \int_0^\infty  \hbar \omega \aopd_\mathrm{in}(\omega) \aop_\mathrm{in}(\omega) \dd \omega$ \cite{Scully_Zubairy_1997}, where $\aop_\mathrm{in}(\omega)$ is a frequency-domain annihilation operator of the field satisfying $\comm{\aop_\mathrm{in}(\omega)}{\aopd_\mathrm{in}(\omega')}=\delta(\omega-\omega')$. The free TLS Hamiltonian is $\Ham_{TLS} = \hbar \omega_0 \sigp \sigm $, where $\sigp = \ket{e}\bra{g}$ (resp. $\sigm = \ket{g}\bra{e}$) is a raising (resp. lowering) operator acting on the TLS.
In the rotating wave and flat coupling approximations, the interaction between the TLS and the field is governed by the Hamiltonian $\Vop = i\hbar \int \sqrt{\frac{\gamma}{2\pi}}\Bigl(\aopd_\mathrm{in}(\omega) \sigm - \aop_\mathrm{in}(\omega) \sigp \Bigr) \dd \omega$ \cite{prasad_closing_2024}, with $\gamma$ the coupling rate of the TLS to the waveguide.
In the interaction picture, $\Vop$ becomes $ \Vop(t) =i\hbar \sqrt{\gamma} \Bigl(\aopd_{\text{in}}(t)\sigm - \aop_{\text{in}}(t) \sigp \Bigr)$ \cite{gardiner_input_1985}, where $\aop_\text{in}(t) = \frac{1}{\sqrt{2 \pi}} \int \dd \omega \aop_\mathrm{in}(\omega) \e^{-i (\omega - \omega_0) t}$ is a time-domain annihilation operator satisfying $\comm{\aop_\mathrm{in}(t)}{\aopd_\mathrm{in}(t')} = \delta(t - t')$. 
The output field is related to the input field and state of the TLS by the input-output relationship $\aop_{\text{out}}(t) = \aop_{\text{in}}(t) + \sqrt{\gamma} \sigm$ \cite{gardiner_input_1985}.

\begin{figure}[t]
    \centering
    \includegraphics[width=0.95\linewidth]{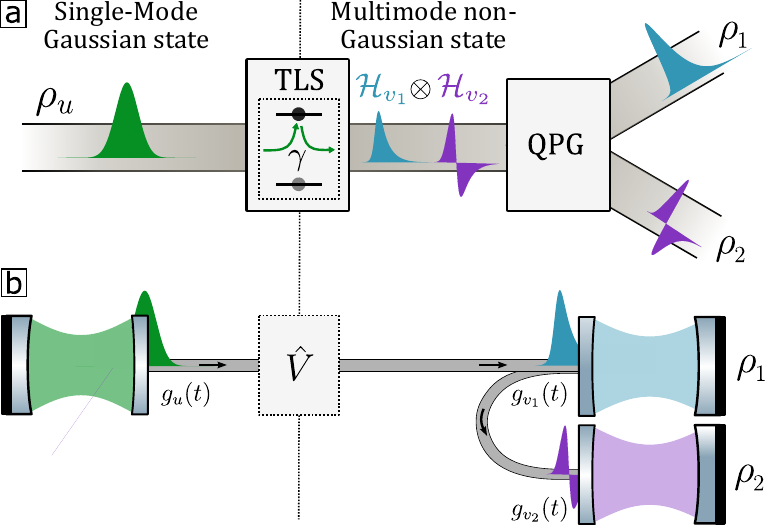}
    \caption{(a) Scheme of the analyzed protocol. A propagating pulsed Gaussian state $\rho_u$, initially occupying a single spectro-temporal mode, interacts with a two-level system and is transformed into a temporally multimode non-Gaussian quantum state. After interacting with the TLS (with a coupling $\gamma$), several modes of the field are populated. In the two-mode regime considered here, the output state supported on $v_1\otimes v_2$ can be mapped onto distinct spatial modes, for instance using a quantum pulse gate (QPG) \cite{Brecht2014}.
    (b) Equivalent diagram of the virtual cavity formalism \cite{kiilerich_input-output_2019, kiilerich_quantum_2020} applied to our two mode system. The input field is initially contained in the input cavity. Thanks to a tailored coupling $g_u(t)$, it leaks a pulse of shape $u(t)$ in the waveguide. The first output cavity captures the field in mode $v_1$ and reflects the rest, and the second output cavity captures the field in mode $v_2$ and reflects the rest, thanks to tailored couplings $g_{v_i}(t)$. }
    \label{fig:virtual_cavities}
\end{figure}

As depicted in Fig. \ref{fig:virtual_cavities}(b), we use the virtual cavity formalism introduced by Kiilerich and Mølmer \cite{kiilerich_input-output_2019, kiilerich_quantum_2020} to describe pulsed interactions between the field and the TLS. In this framework, the input and output temporal modes are mathematically represented by virtual cavities with time-dependent coupling rates.
We consider an input field described by a single spectral-temporal mode of envelope $u(t)$ such that $\int \abs{u(t)}^2 \dd t = 1$, with an associated bosonic operator $\aopd_u = \int \dd t u(t) \aopd_\text{in}(t)$ populated by an input state $\ket{\psi}_{\text{in}}$. The pulse is emitted from an input virtual cavity through a tailored coupling $g_u(t)$. 
After its interaction with the TLS, the output field generally occupies several temporal modes. A natural modal basis can be obtained by diagonalizing the output-field correlation function \cite{kiilerich_input-output_2019,kiilerich_quantum_2020}.

In this work, we focus on input pulses of average photon number $\langle n \rangle$ and characteristic duration $\tau$ such that $\langle n \rangle \gamma \tau \lesssim 1$. In this parameter regime,
the two dominant modes contain nearly all of the output population (See Fig. \ref{fig:bimodality} in Appendix \ref{supp:virtual_beamsplitter}).
This basis, however, is not unique: different orthonormal mode decompositions can reveal different properties of the output state. In this work, we exploit this freedom to analyze and optimize the modal structure of the generated non-Gaussian state. This perspective is closely related to recent work on the temporal-mode structure of few-photon states scattered by a two-level system \cite{Bouchereau2026}.
Let's call $\mathcal{V}$ the subspace of spectral-temporal modes generated by these two modes. 
We choose an orthonormal basis ($v_1$, $v_2$) of $\mathcal{V}$ (coupled to two virtual cavities with rates $g_{v_1}(t)$, $g_{v_2}(t)$ as shown in fig \ref{fig:virtual_cavities}). 
Calling $\mathcal{H}_{v_i}$ the Hilbert space of states in mode $v_i$, we can compute numerically the joint density matrix $\rho_{1,2}$ of the output state in $\mathcal{H}_{v_1} \otimes \mathcal{H}_{v_2}$ at the end of the interaction between the input pulse and the TLS. 
We call $\rho_{i}$ the state of the field projected on $v_i$.
Although $v_1$ and $v_2$ may be chosen as the two dominantly populated eigenmodes, any orthonormal basis of $\mathcal V$ can be obtained through a linear mode rotation
$   \begin{pmatrix}
         v_1^{\theta} \\
         v_2^{\theta}
     \end{pmatrix}
     =
     \begin{bmatrix}
         \cos(\theta) & \sin(\theta) \\
         - \sin(\theta) & \cos(\theta)
     \end{bmatrix}
     \begin{pmatrix}
         v_1 \\
         v_2
     \end{pmatrix}$.
This mathematical freedom allows us, for example, to choose the output-mode basis that maximizes the purity of the reduced states $\rho_1$ and $\rho_2$. 
The corresponding transformation is equivalent to applying a virtual beam splitter operator to $\rho_{1,2}$ and is detailed in Appendix \ref{supp:virtual_beamsplitter}. 
It allows us to explore different output-mode decompositions without solving the full emitter-field dynamics every time.

\section{Deterministic generation of Non gaussian states with coherent pulsed drive} \label{sec:coherent_input}

We now consider a coherent pulse occupying a single temporal mode with Gaussian envelope of characteristic duration $\tau$, centered around time $t_p$: $u(t) \propto \exp(-\frac{(t - t_p)^2}{2 \tau^2})$. The input coherent state is $|\psi\rangle_{\mathrm{in}} = \D_u(\alpha) \ket{0} = \ket{\alpha}_u$, where $\D_u(\alpha) = \exp(\alpha \aopd_u - \alpha^* \aop_u)$. The pulse is characterized by its duration $\tau$ and mean photon number $\expval{n} = \abs{\alpha}^2$. We take $\alpha \in \mathcal{R}^+$ for simplicity. 

\subsection{Choice of output spectro-temporal modes}
After identifying the two dominantly populated output modes, we optimize their linear combination to maximize the purity of the corresponding reduced states $\rho_{1}$ and $\rho_{2}$. When the resulting joint state satisfies $\rho_{1,2}\simeq\rho_1\otimes\rho_2$, the state in either mode can be generated deterministically, without conditioning on a measurement of the other mode. The optimization procedure is detailed in Appendix \ref{supp:virtual_beamsplitter}.

Fig. \ref{fig:coherent_input} presents several properties of $\rho_1$ as functions of $|\alpha|^2\in[0.1,2.0]$ and $\gamma\tau\in[0.1,2.0]$. Fig.~\ref{fig:coherent_input}(a) shows its impurity, $1 - \Tr{\rho_{1}^2}$. 
The purity decreases as the product $|\alpha|^2\gamma\tau$ increases, reflecting the increasing population of additional temporal modes. Nevertheless, for $|\alpha|^2\gamma\tau\lesssim0.5$, the purity of $\rho_1$ remains above $0.99$.
The purity is also maximal in the limits of very short pulses or vanishing input amplitude. These limits are, however, physically trivial: either the pulse is too short to interact efficiently with the emitter, or the input field approaches vacuum. The relevant regime is therefore centered around pulse durations comparable to the emitter lifetime ($\gamma \tau \lesssim 1)$.

\subsection{Generation of non-Gaussian states with stellar rank 1}
\begin{figure*}[t]
    \centering
    \includegraphics[width=0.85\linewidth]{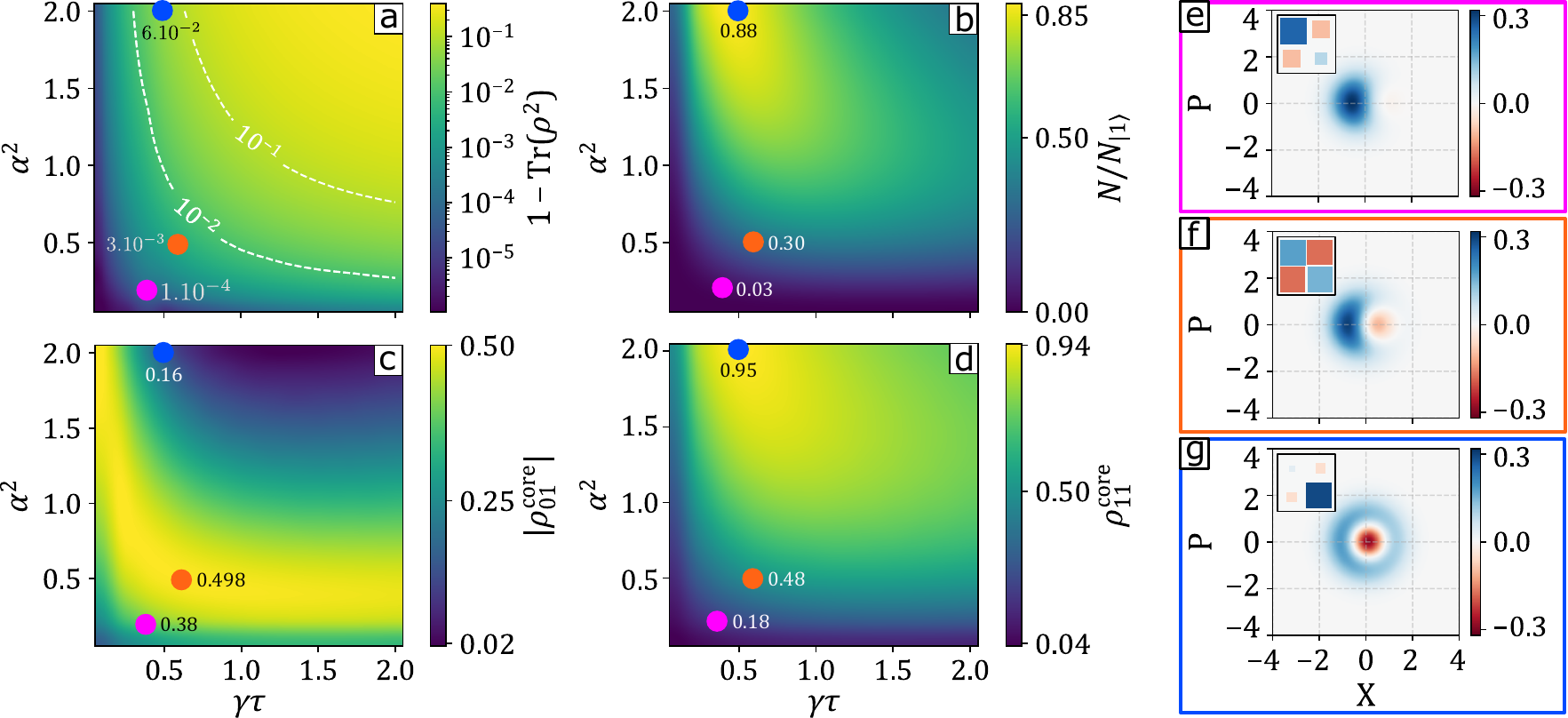}
    \caption{Maps of selected properties of $\rho_{1}$ as functions of the coherent input pulse intensity $|\alpha|^2$ and normalized duration $\gamma\tau $. (a) Impurity $1-\mathrm{Tr}(\rho_1^2)$ of the first output mode $\rho_{1}$. (b) Wigner negativity volume of $\rho_{1}$ normalized to that of a single-photon. (c) Coherence $|\rho_{01}^\mathrm{core}|=|\bra{0}\rho_{1}^{\mathrm{core}}\ket{1}|$ of the corresponding core state. (d) Single photon population $\rho_{11}^\mathrm{core}=\bra{1}\rho_{1}^{\mathrm{core}}\ket{1}$ of the core state. (e-g) Wigner functions of $\rho_{1}$ for three representative values of $\alpha^2$ and $\gamma\tau$. The insets shows the density matrices of the corresponding core states $\rho_{1}^\text{core}$. For similar pulse durations, higher values of $\abs{\alpha}^2$ lead to higher values of $\rho_{11}^\mathrm{core}$. The color of each frame matches the associated marker in panels (a)-(d).}
    \label{fig:coherent_input}
\end{figure*}

We next characterize the states generated in the optimized output mode. Fig.~\ref{fig:coherent_input}(b) shows the Wigner negativity volume $\mathcal{N}$ of $\rho_1$, normalized by that of a single-photon state (see definition in Appendix \ref{supp:definitions}). 
The largest negativity is obtained for short and relatively intense pulses ($\alpha^2 = 2$, $\tau = 0.5 \gamma^{-1}$), consistent with a near-$\pi$-pulse excitation of the emitter. Here we reach $\mathcal{N} / \mathcal{N}_{\ket{1}} = 0.88$ (blue dot in Fig. \ref{fig:coherent_input}), but we can get arbitrarily close to $1$ by further increasing $\alpha$, as shown in \cite{Lim2026}, where energetic cost functions are introduced to identify output modes carrying large Wigner negativities.

To characterize the genuinely non-Gaussian structure of the generated states, we use the notion of core state \cite{Menzies2009,Lachman2019}. Indeed, a state can have a large photon number simply because of Gaussian displacement or squeezing, without involving a more complex non-Gaussian resource. By removing such Gaussian contributions, the core state reveals the minimal Fock-space structure underlying the non-Gaussian part of the state. This notion is closely related to non-Gaussian complexity measurements such as the stellar rank \cite{Fiurasek2022, Chabaud2023}.
For each pair of parameters $\expval{n} = \abs{\alpha}^2$ and $\gamma\tau$, we determine the core state by removing the displacement component of $\rho_i$. We apply an adjustable displacement $\D(\beta_i)$ and define $\rho_i^{\mathrm{core}}(\beta_i)=\D(\beta_i)\rho_i\D^\dag(\beta_i)$. The optimal value $\beta_i^\star$ is chosen so as to make the photon-number distribution of $\rho_i^{\mathrm{core}}(\beta_i)$ as concentrated as possible in the lowest Fock levels, thereby minimizing the number of significantly populated Fock states. We then define $\rho_i^{\mathrm{core}}=\rho_i^{\mathrm{core}}(\beta_i^\star)$. This procedure yields a displacement-free representation of the state and allows us to identify the minimal core Fock subspace carrying its non-Gaussian component.

We find that, throughout the considered parameter range, the core state has negligible population above the single-photon level and is therefore effectively confined to the subspace spanned by $\{|0\rangle,|1\rangle\}$. This is consistent with the origin of the nonlinearity: a single two-level emitter saturates at one excitation and therefore primarily modifies the vacuum and single-photon components of the core state.
Fig.~\ref{fig:coherent_input}(c) shows the matrix element $(\rho_{1}^{\text{core}})_{01}=\bra{0}\rho_{1}^{\text{core}}\ket{1}$, while Fig.~\ref{fig:coherent_input}(d) shows the matrix element $(\rho_{1}^{\text{core}})_{11}$.
These two maps show that driving a TLS with a coherent pulse can generate displaced superpositions of $\ket{0}$ and $\ket{1}$ with arbitrary weights.
To illustrate this, Fig.~\ref{fig:coherent_input}(e-g) shows the Wigner functions of the output states $\rho_{1}$ and the density matrices of their corresponding core states $\rho_{1}^{\text{core}}$ for three representative sets of parameters. The positions of these states are indicated on the different color maps by colored dots, highlighting the corresponding regimes.
In particular, for $\alpha^2 = 0.5$ and $\tau = 0.6 \gamma^{-1}$ (orange marker), the output core state is well approximated by: $\rho_{1}^\text{core} = \ket{\psi_{v_1}^\text{core}} \bra{\psi_{v_1}^\text{core}}$, with $\ket{\psi_{v_1}^\text{core}} = \frac{1}{\sqrt{2}}(\ket{0} - \ket{1})$. Here we have a purity of $0.997$ and $|(\rho_{1}^{core})_{01}|=0.498$, thus deterministically overcoming a noteworthy threshold for quantum non-Gaussian coherence \cite{LachmanCoherences}. We can get even closer to 0.5 by fine tuning $\alpha$ and $\tau$ around those values. Moreover, for an average of $2$ input photons ($\abs{\alpha}^2 = 2$), the output core state already exhibits a single-photon probability of $0.95$ (Fig. \ref{fig:coherent_input}(g)). This surpasses an asymptotic multicopy threshold for photon bunching, opening the possibility of breeding complex non-Gaussian states \cite{Zapletal_Filip_2017, Zapletal:21}.

Overall, coherent pulses can be used to generate high-purity non-Gaussian states deterministically. For $\langle n \rangle \gamma \tau \lesssim 1$, the resulting states are limited to displaced superpositions of $\ket{0}$ and $\ket{1}$, with at most a single negative region in the Wigner function and a stellar rank of $1$. 
To go beyond these limitations, and following a strategy similar to that proposed in \cite{leonhardt_wigner-negative_2025}, we turn to a squeezed input to test single photon emitter for higher Fock states in the core.

\section{Higher-rank Non-Gaussian states under a squeezed-vacuum input pulse } \label{sec:squeezed_input}

To go beyond the single-photon core state subspace, we turn to another photonic state which can be routinely and deterministically produced in quantum optics experiments: a squeezed vacuum state.
Pulsed squeezed vacuum can be generated by pumping broadband optical parametric oscillators or nonlinear waveguides \cite{Ast_2013, terrasson2026}. 
To follow the type of temporal input modes typically emitted by cavity-based sources, we choose an input mode with a Lorentzian spectrum, corresponding to a double-exponential envelope: $u(t) \propto \exp(\frac{- \abs{t - t_p}}{\tau})$.
The input state is squeezed vacuum $\ket{\psi}_\text{in} = \hat{S}_u(z) \ket{0}$, where z is the squeezing parameter, and $\hat{S}_u(z) = \exp(\frac{1}{2}\left(z^* {\aop_u}^2 - z {\aopd_u}^2\right))$ the squeezing operator.
As in the coherent-input case, we analyze the modal decomposition of the output field and restrict our study to a regime in which the two dominant modes contain more than $99\% $ of the output population.
We'll focus here also on one set of parameters that provides particularly interesting results: $8$dB of squeezing and $\gamma \tau = 0.25$. There, the input pulse carries an average of $\langle n \rangle = 1.15$ photons.
In this regime, $v_1$ and $v_2$, the two most populated output modes, represent $99.7 \%$ of the total output population.

The corresponding joint density matrix $\rho_{1, 2}$ has itself a purity of $99.5 \%$. In the following, to give a clearer picture of the output state, we will therefore approximate it as a pure state $\ket{\Psi}_{1, 2}$. 

\subsection{Decomposition into even- and odd- parity components}

The input squeezed vacuum state is by definition a superposition of even Fock states, and in the absence of loss or noise, the total photon number is conserved. Consequently, the photon numbers in output modes $v_1$ and $v_2$ must have the same parity: an even (odd) photon number in one mode is necessarily accompanied by an even (odd) photon number in the other. Hence the joint output state can be decomposed as 
\begin{equation}
\begin{aligned}
\ket{\Psi}_{1, 2} &= \sum_i a_i \ket{\psi_i^\text{even}}_1 \otimes \ket{\phi_i^\text{even}}_2+ \\
&\phantom{={}}\sum_j b_j\ket{\psi_j^\text{odd}}_1 \otimes \ket{\phi_j^\text{odd}}_{2} \\
&=a \ket{\Psi^\text{even}}_{1, 2} + b \ket{\Psi^\text{odd}}_{1, 2}
\end{aligned}
\end{equation}
This decomposition into parity components holds for any orthogonal basis ($v_1^\theta$, $v_2^\theta$) of $\mathcal{V}$. 
Let's consider again two orthogonal modes upon the most populated mode basis that we can choose.
By tracing out the other still populated mode, the reduced state in mode $v_i^\theta$ can therefore be written as:
\begin{equation}
\rho_{i,\theta} = \abs{a}^2 \rho^{\text{even}}_{i,\theta} + \abs{b}^2 \rho_{i,\theta}^{\text{odd}}
\end{equation}
The conditional states $\rho_{i,\theta}^{\mathrm{even}}$ and $\rho_{i,\theta}^{\mathrm{odd}}$ are not necessarily pure.
This is illustrated in Fig. \ref{fig:squeezed_input}(a), which shows the purities of $\rho_{1, \theta}$, $\rho_{1,\theta}^{\text{even}}$ and $\rho_{1,\theta}^{\text{odd}}$ as functions of $\theta$. 

In the following, and in Fig. \ref{fig:squeezed_input}, we investigate the output modes states for different values of $\theta$ ($\theta_1$, $\theta_2$, $\theta_3$), resulting on different purity properties marked by different colored dots in Fig. \ref{fig:squeezed_input}(a) (respectively green, pink and blue). The corresponding first output modes $v_1^\theta(t)$ are shown in Fig. \ref{fig:squeezed_input}(b), with the second modes $v_2^\theta(t)$ in the insets, next to the incoming wave-packet shape $u(t)$.
In Fig. \ref{fig:squeezed_input}(c,d,e) is shown the resulting different Wigner functions of the output states $\rho_{1, \theta}$, $\rho_{1,\theta}^{\text{even}}$ and $\rho_{1,\theta}^{\text{odd}}$ and, in the insets the corresponding Wigner functions of $\rho_{2, \theta}$, $\rho_{2,\theta}^{\text{even}}$ and $\rho_{2,\theta}^{\text{odd}}$.

\subsection{Generation of Higher-rank non-Gaussian states}

\begin{figure}[t]
    \centering    \includegraphics[width=1\linewidth]{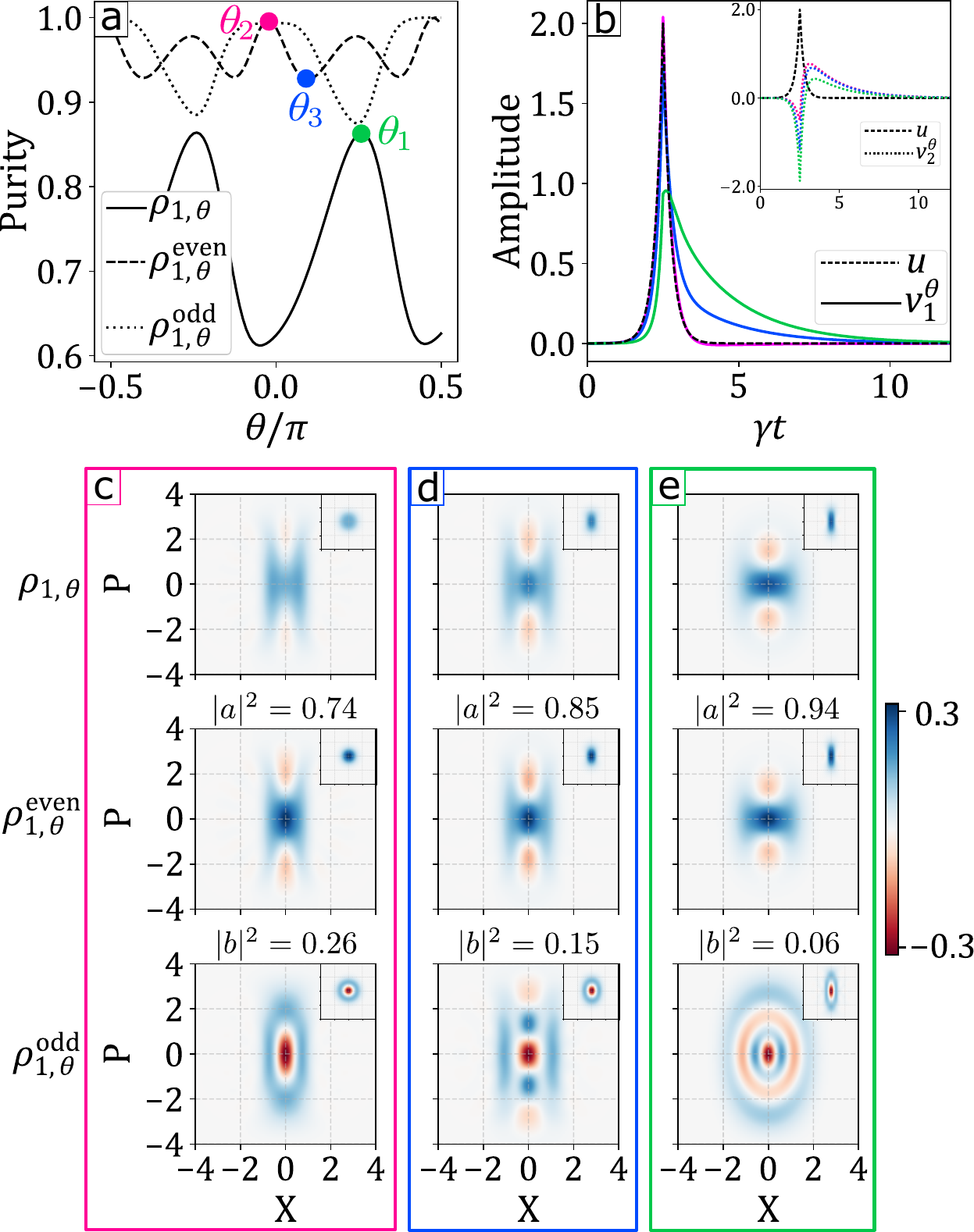}
    \caption{Bi-modal non-Gaussian output from a squeezed vacuum input pulse. (a) Purity of $\rho_{1,\theta}$, $\rho_{1, \theta}^\text{even}$ and $\rho_{1, \theta}^\text{odd}$ as a function of $\theta$. Three values of $\theta$ are marked by colored dots: $\theta_1 =0.26 \pi$, $\theta_2 = -0.02 \pi$ and $\theta_3 = 0.09\pi$. (b) Output modes $v_1^\theta$ (full) and $v_2^\theta$ (dashed) for $\theta_{1,2,3}$. (c-e) Output states for $\theta_{1,2,3}$. From top to bottom, the Wigner functions correspond to the density matrices $\rho_{1,\theta}$, $\rho_{1, \theta}^\text{even}$ and $\rho_{1, \theta}^\text{odd}$. The insets correspond to the matching density matrices in $v_2^\theta$: $\rho_{2,\theta}$, $\rho_{2, \theta}^\text{even}$ and $\rho_{2, \theta}^\text{odd}$. For each column and $i \in \{1,2\}$, $\abs{a}^2$ and $\abs{b}^2$ are such that $\rho_{i,\theta} = \abs{a}^2 \rho^{\text{even}}_{i,\theta} + \abs{b}^2 \rho_{i,\theta}^{\text{odd}}$.}
    \label{fig:squeezed_input}
\end{figure}

We first consider the value of $\theta$ that maximizes the purity of $\rho_{1, \theta}$, corresponding to the green point in Fig. \ref{fig:squeezed_input}(a) and to $\theta_1 = 0.26\pi$. Fig.~\ref{fig:squeezed_input}(b) shows the corresponding output mode $v_1^{\theta_1}$, and Fig.~\ref{fig:squeezed_input}(e) shows the Wigner functions of the output state and of its even- and odd-parity components.
The resulting state $\rho_{1,\theta_1}$ already has a relatively high purity at $0.86$, and closely resembles a squeezed even cat state. This shows that a relatively pure and strongly non-Gaussian state, with a core state extending up to the two-photon level, can be generated deterministically. 

The parity decomposition further reveals that an even purer squeezed cat state is included in the mixture. Indeed, $\rho_{1,\theta_1}$ is dominated by its even component, with weight $|a|^2=0.94$. 
The state $\rho_{1,\theta_1}^{\mathrm{even}}$ has a fidelity of $99.4\%$ with a squeezed even cat state with $z=4$ dB and $\alpha^2=1$, and a purity of 0.98. 
The odd component has a smaller weight, $|b|^2=0.06$, and $\rho_{1,\theta_1}^{\mathrm{odd}}$ resembles a squeezed three-photon state, with $95\%$ fidelity for $z=1.6$ dB, although with a lower purity ($0.88$).
In the second mode, $\rho_{2,\theta_1}^{\mathrm{odd}}$ resembles a squeezed single-photon state, suggesting an effective transfer of one photon from mode 1 to mode 2. Fidelity-based stellar rank witnesses guarantee a rank $r^* \geq 2$ for $\rho_{1,\theta_1}^{\mathrm{even}}$ and $r^* \geq 3$ for $\rho_{1,\theta_1}^{\mathrm{odd}}$ (see Appendix \ref{supp:stellar_rank}).

To obtain purer heralded non-Gaussian states, one can instead directly maximize the purities of the parity-resolved components $\rho_{1,\theta}^{\mathrm{even}}$ and $\rho_{1,\theta}^{\mathrm{odd}}$.
The pink point in Fig.~\ref{fig:squeezed_input}(a), corresponding to $\theta_2=-0.02\pi$, marks a simultaneous maximum of both purities.  Interestingly, for this value of $\theta$, the temporal profile $v_1^{\theta_2}(t)$ closely matches that of the incoming wave-packet $u(t)$ (fig~\ref{fig:squeezed_input}(b)). The corrresponding output state Wigner functions are shown in Fig.\ref{fig:squeezed_input}(c).
Although the full state $\rho_{1,\theta_2}$ exhibits little Wigner negativity, this mode basis still gives access to interesting conditional states. The odd component has a weight $|b|^2=0.26$, and $\rho_{1,\theta_2}^{\mathrm{odd}}$ closely resembles a squeezed single-photon state, with $99.6\%$ fidelity for $4.2$ dB of squeezing, while $\rho_{2,\theta_2}^{\mathrm{odd}}=\ket{1}\bra{1}$. This process can therefore be interpreted as the removal of a single photon from the input pulse into the mode $v_2^{\theta_2}$, similarly to Ref. \cite{lund_subtraction_2024}.
Additionally, $\rho_{1, \theta_2}^\text{even}$ has a fidelity of $99.8\%$ with a squeezed even cat state $\ket{\phi} \propto S(z) ( \ket{\alpha} + \ket{-\alpha}) $, with $z=7$dB and $|\alpha|^2=1$, with vacuum in the second mode: $\rho_{2,\theta_2}^\text{even} = \ket{0}\bra{0}$. Both $\rho_{1, \theta_2}^\text{even}$ and $\rho_{1, \theta_2}^\text{odd}$ have a purity above $0.99$.
Thus, in the output mode $v_1^{\theta_2}$, the pulse has a $74\%$ probability of generating a squeezed even cat state and a $26\%$ probability of generating a squeezed single photon. A parity measurement on mode $v_2^{\theta_2}$  can reveal which process happened. For $\theta_2$, fidelity-based stellar rank witnesses guarantee a rank $r^* \geq 2$ for $\rho_{1, \theta_2}^\text{even}$ and a rank $r^* \geq 1$ for $\rho_{1, \theta_2}^\text{odd}$.

\begin{figure*}[t]
    \centering
    \includegraphics[width=1.0\linewidth]{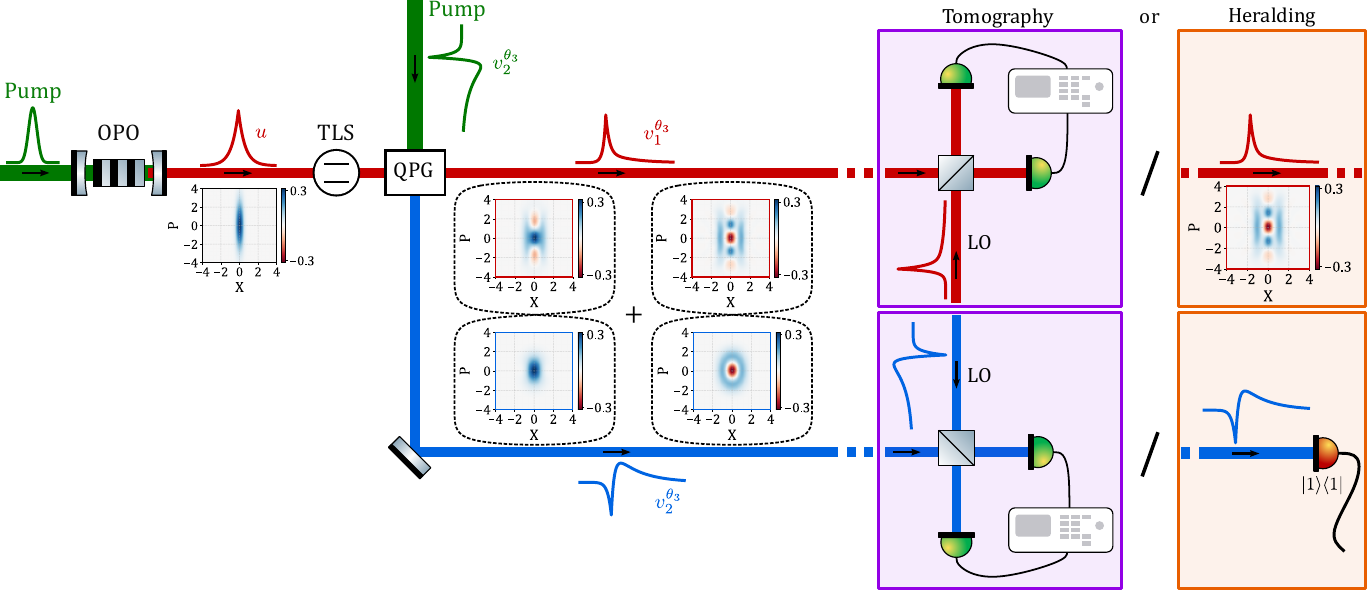}
    \caption{Simplified experimental protocol. A coherent pulse is sent on an optical parametric oscillator (OPO) to generate a squeezed vacuum pulse. The interaction with the TLS generates a bi-modal output field with Non-Gaussian features. A quantum pulse gate separates modes $v_1^\theta$ and $v_2^\theta$. One can then perform tomography on the bimodal output field or implement a heralding protocol. The dashed lines around the Wigner functions represent entanglement between the modes. Here we have chosen to represent the protocol with the output mode basis ($v_1^{\theta_3}$, $v_2^{\theta_3}$).}
    \label{fig:heralding_protocol}
\end{figure*}

We further explore this heralding strategy by considering another output-mode basis.  The blue point shown in Fig.~\ref{fig:squeezed_input}(a), corresponding to $\theta_3=0.09\pi$, marks a local minimum in the purity of $\rho_{1,\theta_3}^{\mathrm{even}}$ (at 0.93), while the purity of $\rho_{1,\theta_3}^{\mathrm{odd}}$ remains high (0.98).
Fig. \ref{fig:squeezed_input}(d) shows the corresponding Wigner functions of the output states. The even component $\rho_{1, \theta}^\text{even}$ still resembles a squeezed even cat with $98.0 \%$ fidelity (with $z=6.5$ dB, $\alpha^2 = 1.2$). 
More strikingly, the odd component $\rho_{1, \theta_3}^\text{odd}$ has a fidelity of $99.3 \%$ with a relatively large squeezed odd cat state $\ket{\phi} \propto S(z) ( \ket{\alpha} - \ket{-\alpha}) $ with $z = 8$ dB and $\alpha^2 = 3.5$, and three negativities present in the momentum. Its weight and therefore its generation probability is $\abs{b}^2 = 0.15$.
In mode $v_2^{\theta_3}$, the even and odd components exhibit a small amount of squeezing but remain predominantly vacuum and single-photon states, respectively. 
For $\theta_3$, stellar rank witnesses guarantee a rank $r^* \geq 2$ for $\rho_{1, \theta_3}^\text{even}$ and a rank $r^* \geq 3$ for $\rho_{1, \theta_3}^\text{odd}$.

\section{Proposed protocol and experimental considerations}

Using a squeezed drive, the interaction generates a bimodal non-Gaussian state that may already constitute a useful resource. To characterize these predictions experimentally, one could use quantum pulse gates \cite{Eckstein:11, SerinoMQPG2023, Brecht2014, Christiansen2026} to project and convert a chosen pair of spectro-temporal modes into distinct frequency modes. These modes could then be separated into different spatial paths, for instance with a dichroic mirror, and characterized by two-mode homodyne tomography. Such a protocol is illustrated in Fig. \ref{fig:heralding_protocol}. In the following, we focus on the deterministic and heralded generation of non-Gaussian states in a single output mode.
For example, selecting mode $v_1^{\theta_1}(t)$ deterministically produces a squeezed even cat state with $\alpha=1$ and purity $0.86$.
Beyond this deterministic regime, the parity structure of the joint state enables the heralding of a conditional state in mode $v_1^\theta(t)$ through a parity measurement on mode $v_2^\theta(t)$.
In particular, for the mode basis associated to $\theta_3$, detecting an odd number of photons on mode $v_2^{\theta_3}(t)$ heralds a large squeezed odd cat state of size $|\alpha|^2 = 3.5$, with a probability of $15\%$. 
A simpler photon-number-resolved measurement can also be used. Projecting mode $v_2^{\theta_3}$ onto $\ket{1}\bra{1}$ generates a similar squeezed cat state, with $|\alpha|^2=3.1$ and $z=7.3$ dB, with a heralding probability of $14\%$, a purity of $0.987$, and a fidelity of $99.4\%$. 
Such large cat states, albeit heralded, could already constitute useful resources for GKP-state generation \cite{Konno_2024}. Indeed, in contrast to OPO/OPA-based large cat-generation protocols that rely on multiphoton detection, or homodyne-conditioned breeding \cite{Huang2015,Etesse2015,Sychev2017,Endo2026}, this protocol starts from a deterministically produced Gaussian squeezed input and requires only a single-photon heralding event. In this respect, it shares a similar practical advantage as the atom-light protocol of Ref.~\cite{Hacker2019}, although the underlying physical mechanism is different.

\subsection{Imperfect coupling and non-chirality}
We now address an important experimental challenge in the generation of the non-Gaussian states discussed above. Realistic waveguide-QED implementations are inevitably subject to loss. A major contribution arises from the coupling of the TLS to unwanted spatial modes, which we collectively describe as a loss channel. For instance, an emitter may couple both to the guided mode and to free-space modes, with any emission into free space being lost. 
We consider for this the simple case where the TLS is excited by a coherent pulse of intensity $\alpha^2 = 0.5$ and duration $\tau = 0.6 \gamma_\text{tot}^{-1}$. 
The TLS remains coupled to the right-propagating waveguide mode at rate $\gamma$, while coupling to the loss channel at rate $\gamma_{\mathrm{loss}}$.
We define the total decay rate as $\gamma_{\mathrm{tot}}=\gamma+\gamma_{\mathrm{loss}}$ and model the loss channel through the Lindblad jump operator $L = \sqrt{\gamma_\text{loss}} \sigma_-$.

\begin{figure}[t]
    \centering
    \includegraphics[width=1\linewidth]{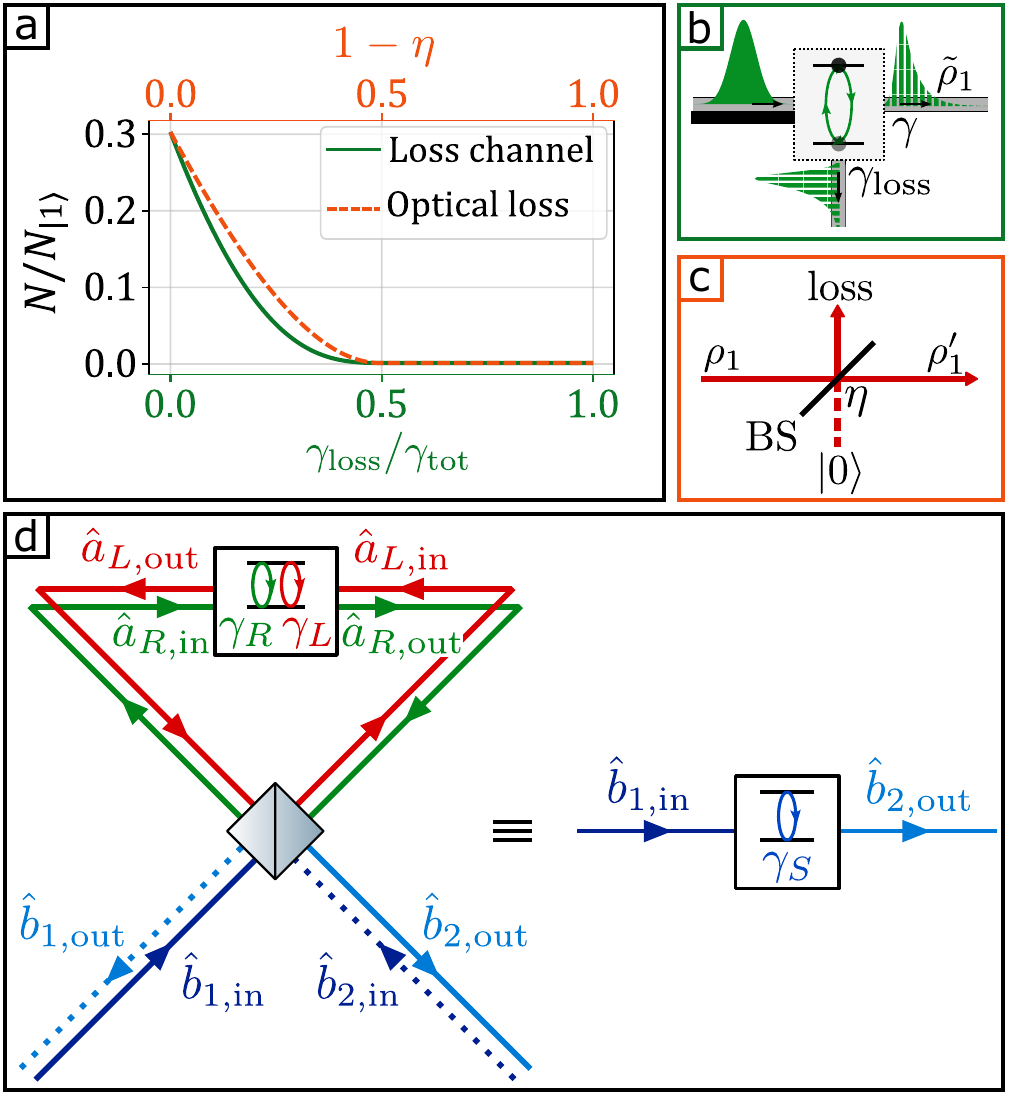}
    \caption{(a) Evolution of the Wigner negativity of the output state as a function of losses, for two different loss processes. The input is a coherent state ($\alpha = \sqrt{0.5}$, $\tau = 0.6 \gamma^{-1}$). (b) Coupling to a loss channel: part of the light emitted by the TLS escapes in another spatial mode. (c) Optical loss: the output state is mixed with vacuum via a beam-splitter of transmittance $\eta$. (d) Set-up to recover an effective chiral system from a partially chiral system. }
    \label{fig:losses}
\end{figure}

Fig. \ref{fig:losses}(a) shows the effect of such imperfect coupling on the essential Wigner negativity of the output state. We plot the normalized Wigner negativity as $\gamma_\text{loss} / \gamma_\text{tot}$ varies between $0$ and $1$ (solid green curve). For comparison, we show the effect of conventional optical losses (dotted orange curve). The two mechanisms are represented in Fig. \ref{fig:losses}(b) and Fig. \ref{fig:losses}(c), respectively. In both cases, the Wigner negativity vanishes at $50 \%$ efficiency. However, imperfect emitter-waveguide coupling is not directly equivalent to applying optical losses: the Wigner negativity drops faster for emission into the loss channel than for regular optical losses. 
This point is particularly important: even when near-unity $\beta$ factors are achieved \cite{Arcari2014}, most platforms remain non-chiral or only partially chiral, so that the emitter couples to both right- and left-propagating modes. If only one propagation direction is collected, emission into the opposite direction is effectively lost. In the following, we propose an experimental configuration that restores an effectively chiral interaction, as illustrated in Fig.~\ref{fig:losses}(d).

\subsection{Recovering an effectively chiral system}
In this section, we show that a bidirectional configuration can be transformed into an effectively unidirectional, chiral one by placing the emitter in a self-interfering geometry, in the spirit of related work on cavity systems \cite{lahadInducedCavitiesPhotonic2017}. 
We now consider an emitter coupled to an ideal lossless bidirectional waveguide, supporting both left- and right-propagating modes. The interaction Hamiltonian then becomes:
\begin{equation}
    \begin{split}
        \hat{V}(t) = i\hbar \sqrt{\gamma_R} \left(\aopd_{R, \text{in}}(t) \sigm - \aop_{R, \text{in}}(t) \sigp \right) \\
        + i\hbar \sqrt{\gamma_L} \left(\aopd_{L, \text{in}}(t) \sigm - \aop_{L, \text{in}}(t) \sigp \right)
    \end{split}
\end{equation}
Where $\aopd_{R, \text{in}}(t)$ (resp. $\aopd_{L, \text{in}}(t))$ is the time-domain creation operator associated to the right-propagating (resp. left-propagating) mode.
We obtain 2 input-output relationships:
\begin{equation}
    \begin{aligned}
        \aop_{R, \text{out}} (t) &= \aop_{R, \text{in}} + \frac{i}{\hbar} \comm{\hat{V}(t)}{\aop_{R, \text{in}}(t)}\\
        &=  \aop_{R, \text{in}} (t) + \sqrt{\gamma_R}\sigm (t) \\
        \aop_{L, \text{out}} (t) &= \aop_{L, \text{in}} + \frac{i}{\hbar} \comm{\hat{V}(t)}{\aop_{L, \text{in}}(t)} \\
        &=  \aop_{L, \text{in}} (t) + \sqrt{\gamma_L}\sigm (t) \\
    \end{aligned}
\end{equation}

Now let us add a beam-splitter with parameters $r$ and $t$ to attempt to recombine the outputs, as depicted in Fig. \ref{fig:losses}(d). The input field will be also split and sent on the TLS from both directions. Following the notations of Fig. \ref{fig:losses}(d), we obtain the following input-output relations:
\begin{equation}
    \begin{aligned}
        \bop_{1, out} &= t \aop_{R, \text{out}} + r \aop_{L, \text{out}}  \\
        &= \bop_{2, in} + \bigl(t\sqrt{\gamma_R} + r \sqrt{\gamma_L} \bigr) \sigma_-\\
        \bop_{2, out} &= -r \aop_{R, \text{out}} + t \aop_{L, \text{out}} \\
        &= \bop_{1, in} + \bigl(-r \sqrt{\gamma_R} + t \sqrt{\gamma_L} \bigr) \sigma_-
    \end{aligned}
\end{equation}

In particular, setting $\gamma_S = \gamma_L + \gamma_R$, $t = \sqrt{\gamma_L/\gamma_S}$ and $r = - \sqrt{\gamma_R/\gamma_S}$ yields:

\begin{equation}
    \begin{aligned}
        \bop_{1, \text{out}} &= \bop_{2,\text{in}} \\
        \bop_{2, \text{out}} &= \bop_{1,\text{in}} + \sqrt{\gamma_S} \sigm \\
        \hat{V}(t) &= i \hbar \sqrt{\gamma_S} \left( \bopd_{1, \text{in}}(t) \sigm - \bop_{1, \text{in}}(t) \sigp \right)
    \end{aligned}
\end{equation}
We have therefore recovered an effective chiral system, where $\bop_{1, \text{in}}$ and $\bop_{2, \text{out}}$ follow the usual input-output relation. We also note that the reciprocal transformation can be obtained in the same way, allowing a fully chiral system to be mapped onto an effective bidirectional platform.

This scheme is straightforward to implement, provided that the relative phase between the two paths is stabilized, and can be applied to both non-chiral and partially chiral systems. Preserving Wigner negativity therefore only requires an overall emitter-waveguide coupling efficiency above $50\%$. We emphasize, however, that this construction applies to a single emitter. For multiple emitters, repeated interactions and non-Markovian dynamics lead to substantially more complex behavior.

\section{Discussion and outlooks}
In summary, we have shown that a single two-level emitter coupled to a waveguide, whether chiral or effectively chiral, is sufficient to generate deterministic, on-demand non-Gaussian states with Wigner negativity under coherent pulsed excitation. The complexity of these states nevertheless remains limited, with low stellar rank and core states confined to the $\{\ket{0},\ket{1}\}$ subspace. More carefully engineered pulse shapes, or extensions to multilevel or multi-emitter systems, may provide access to higher stellar ranks, albeit at the cost of increased experimental complexity.

Alternatively, using a squeezed-vacuum input opens a different regime, in which the interaction no longer produces a single relevant output mode but a structured two-mode non-Gaussian state. Rather than viewing this multimode structure only as a complication, it can be regarded as a resource in itself, following investigations already done for microwaves \cite{Yang_2025}. Its full characterization, including whether it contains useful forms of non-Gaussian correlations or entanglement, is an interesting direction for future work.  Experimentally, this state could be accessed using quantum pulse gates which have demonstrated efficiencies above $80\%$ \cite{Brecht2014}, both to separate the relevant temporal modes and to perform mode-selective two-mode homodyne tomography.
For longer, narrower-bandwidth pulses, fiber-based frequency conversion based on Bragg-scattering four-wave mixing \cite{Christensen2015} may provide a more natural route toward efficient temporal-mode-selective conversion.

In this work, we focused on a simple use case of this two-mode state: heralding a single-mode non-Gaussian resource by measuring one of the temporal modes. In particular, projection of one output mode onto $\ket{1}\bra{1}$ heralds a large squeezed cat state in the other mode with a probability of approximately $14\%$. Such states are relevant to existing optical bosonic-state engineering protocols, including cat-based routes toward GKP-state preparation \cite{Konno_2024,Kiryu_2026}. This provides a complementary perspective to recent proposals using squeezed few-photon superpositions as building blocks for cat-state generation \cite{Luo_2026}, also with the use of QPG and single photon PNR detection. Here, the input is instead a deterministic Gaussian resource, which can be deterministically and relatively easily produced in a quantum optics lab.
However, and for this protocol to work, beyond the need of an efficient QPG, photon-number-resolving detection of a single photon is required. Although detector speed has long been a major limitation, TES detectors can now operate in the MHz regime \cite{larsen_GKP_2025}, while SNSPD-based PNR detectors combine single-photon detection efficiencies above $85\%$ with operating rates of several tens of MHz \cite{Ding2025}.
Combining these figures would yield a conservative overall heralding probability of $\simeq 10^{-1}$ per generated large cat state, which might help to speed up the preparation of approximate GKP states to the $100$kHz to MHz range.
An additional advantage of the proposed protocol is that it requires only a single two-level emitter. Among candidates, solid-state emitters could therefore be particularly promising candidates, owing to their high decay rates, typically ranging from tens of MHz to the GHz regime,  and consequently their compatibility with high-repetition-rate operation. 
Moreover they can typically reach high coupling efficiencies \cite{Arcari2014}, including in chiral nanophotonic structures \cite{lodahl_chiral_2017}. And although chirality is technically required to avoid losing the reflected component of the interaction, corresponding to an effective $50\%$ loss, this constraint can be relaxed using the interferometric scheme introduced above, at the cost of stabilizing the relative phase between the two paths. Recent progress on foundry-compatible quantum-dot spin--photon interfaces, including high-efficiency on-demand photon generation, near-unity photon indistinguishability, spin--multi-photon entanglement, and direct quadrature reconstruction of Wigner-negative single photons, further supports solid-state emitters as realistic platforms for such protocols \cite{Huet2026}.
Taken together, these considerations suggest that squeezed-cat states and therefore GKP-resource states could be generated at much faster repetition rates, which would represent a substantial improvement over current probabilistic approaches.
It also remains an open question whether the generated two-mode state could itself be used directly as a resource for deterministic GKP breeding \cite{Eaton_2019,Winnel_2024}. More broadly, the approach introduced here could provide a useful tool for the efficient generation and characterization of complex non-Gaussian optical states, in particular as resources for bosonic error-correction codes and fault-tolerant universal quantum computation.

\section*{Acknowledgments}

The authors thank Kian Hwee Lim, Kiarn T. Laverick, Samyak P. Prasad, Maria Maffei and Alexia Auffèves for fruitful exchanges, as well as Yann Bouchereau, Mattia Walschaers and Nicolas Treps. The authors also warmly thank Victor Rueskov Christiansen for making his code publicly available on GitHub. This project is supported by the Plan France 2030 through the projects NISQ2LSQ (Grant ANR-22-PETQ-0006), OQuLus (Grant ANR-22-PETQ-0013). R. F. and G.P.T. acknowledge support from the European Union's Horizon Europe research and innovation program under grant agreement No. 101257526 (SUPERSPIN) and No. 101080173 (CLUSSTEC). We also received support from the project CLUSSTAR (8C2024003) and the project CZ.02.01.010022\_0080004649 (QUEENTEC) of EU and the Czech Ministry of Education, Youth and Sport. Project CLUSSTAR has also received funding from the EU Horizon Programme under Grant Agreement No. 731473 and 101017733 (QuantERA).

\appendix
\section{Virtual beam-splitter} \label{supp:virtual_beamsplitter}

\begin{figure}[t]
    \centering
    \includegraphics[width=1\linewidth]{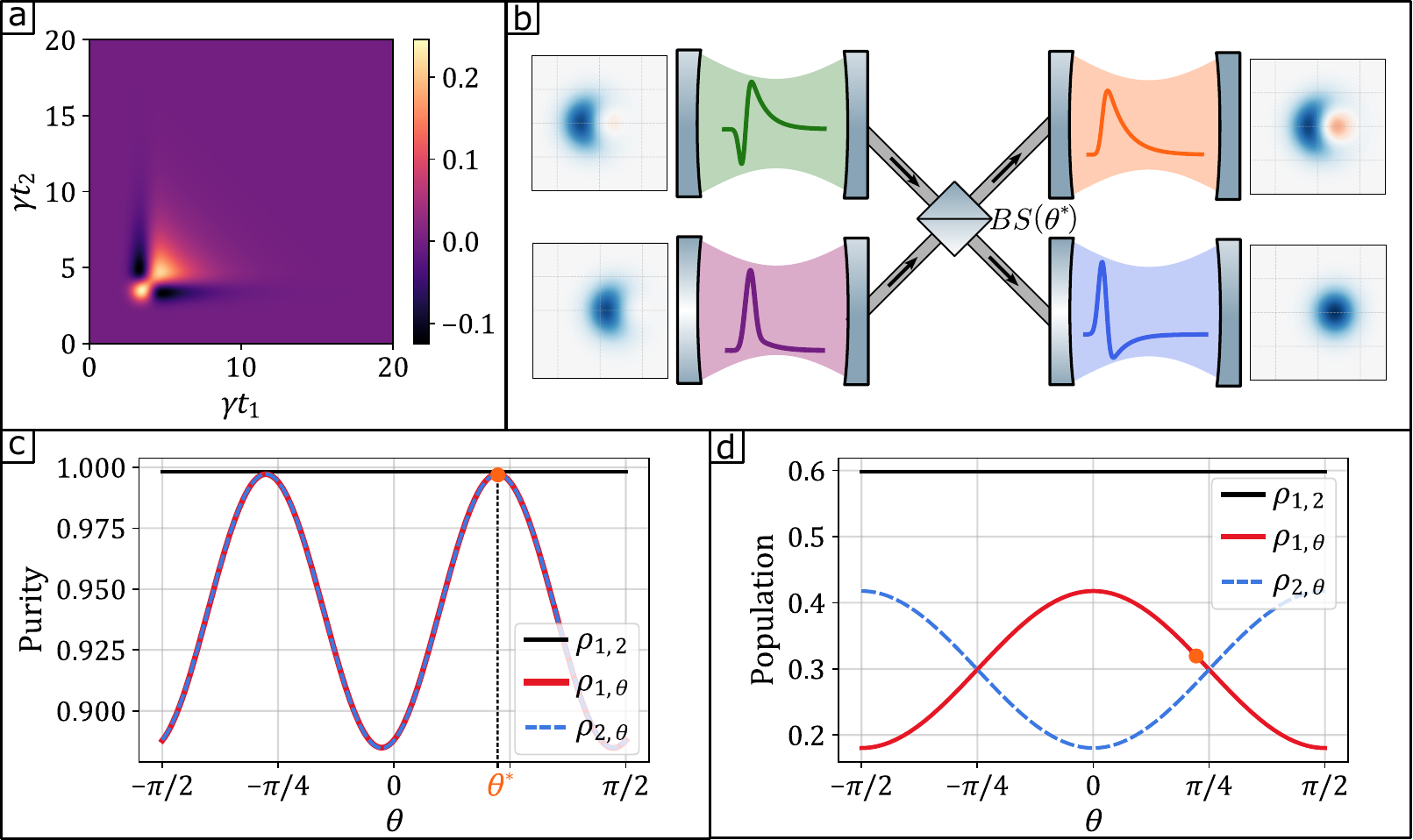}
    \caption{Finding the optimal output modes, example with a coherent input of amplitude $\alpha = \sqrt{0.5}$ and duration $ \gamma \tau = 0.6$ (orange dot in Fig. \ref{fig:coherent_input}). (a) Autocorrelation matrix of the output field $g(t_1, t_2) = \langle \aopd_\text{out}(t_1) \aop_\text{out}(t_2)\rangle$. $v_1$ and $v_2$ are obtained by diagonalizing $g$ and picking the eigenmodes associated to the two largest eigenvalues. (b) Virtual beam-splitter scheme. After solving for the output modes $v_1$ and $v_2$, we perform a beam-splitter operation on $\rho_{1,2}$ to obtain the result for $v_1^\theta$ and $v_2^\theta$ without solving the ME again. The Wigner functions of partial density matrices $\rho_1$, $\rho_2$, $\rho_{1, \theta^*}$ and $\rho_{2, \theta^*}$ are plotted next to the cavities. The mode envelopes are also represented. (c) Purity of the partial density matrices $\rho_{1, \theta}$ and $\rho_{2, \theta}$ as a function of $\theta$. The optimal beam-splitter angle $\theta^*$ is chosen to maximize the purity of $\rho_{1, \theta}$. (d) Population in the modes $v_1^\theta$ and $v_2^\theta$ as a function of $\theta$.}
    \label{fig:VirtualBS}
\end{figure}

In this appendix, we describe our method to choose relevant output modes, as illustrated in fig \ref{fig:VirtualBS}. Following the method from \cite{kiilerich_input-output_2019, kiilerich_quantum_2020}, we compute the autocorrelation matrix of the output field $g(t_1, t_2) = \langle \aopd_\text{out}(t_1) \aop_\text{out}(t_2)\rangle$. Diagonalizing $g(t_1, t_2)$ gives us a set of eigenvalues $\lambda_k$ and eigenmodes $v_k$. $\lambda_k$ is the average population in mode $v_k$. We pick the eigenmodes $v_1$ and $v_2$ (associated to the two largest eigenvalues $\lambda_1$ and $\lambda_2$) and we solve the evolution of the system described in Fig. \ref{fig:virtual_cavities} to  obtain the joint density matrix $\rho_{1,2}$. We then perform a beam-splitter transformation on $\rho_{1,2}$:
$\Tilde{\rho}_{1,2} = \hat{BS}(\theta) \rho_{1,2} \hat{BS}^\dag$, where $\hat{BS}(\theta) = \e^{\theta(\aopd_1 \aop_2 - \aop_1 \aopd_2)}$. We can then vary $\theta$ and optimize for a given quantity, such as the purity of partial density matrices $\rho_{1, \theta}$ and $\rho_{2,\theta}$. These partial density matrices describe the state in output modes $v_1^\theta$ and $v_2^\theta$. Therefore, we only need to solve the evolution of this system for one basis of $\mathcal{V}$, and we can access the state in any mode $v \in \mathcal{V}$. 

\begin{figure}[t]
    \centering
    \includegraphics[width=0.98\linewidth]{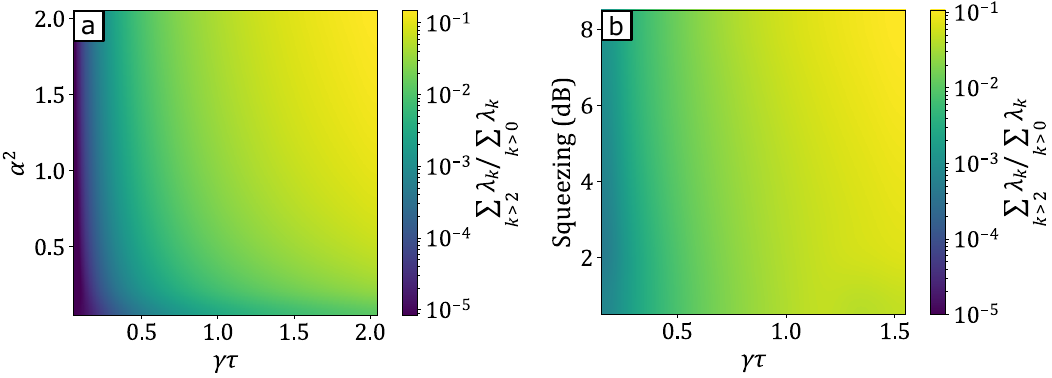}
    \caption{Distance to the bi-modal regime for either a coherent input (a) or a squeezed vacuum input (b). We plot the population in modes $v_{i \geq 3}$, normalized by the total population of the output. For both types of inputs, we see that the duration of the input pulse is a more sensitive parameter than the input population (under these scales).}
    \label{fig:bimodality}
\end{figure}

We have only looked at the two dominantly populated modes and neglected the contribution of other eigenmodes of $g(t_1, t_2)$. In Fig. \ref{fig:bimodality}, we plot the weight of eigenvalues $\lambda_{k \geq 3}$, normalized by the total population of the field, against input pulse parameters. For both coherent and squeezed inputs, we observe that the contribution of these modes is smaller for short input pulses than for long input pulses. The average population of the input pulse also has an effect on the population of these modes, as more photons lead to more modes being populated. 

\begin{table*}[t]
    \centering
    \begin{tabular}{|c|c|c|c|c|c|c|c|c|c|c|c|c|}
        \hline
         &  \multicolumn{4}{|c|}{$\rho_{1, \theta}$} & \multicolumn{4}{|c|}{$\rho_{1, \theta}^\text{even}$} & \multicolumn{4}{|c|}{$\rho_{1, \theta}^\text{odd}$}\\
         \cline{2-13}
         & $F_0^*$ & $F_1^*$ & $F_2^*$ & $F$ & $F_0^*$ & $F_1^*$ & $F_2^*$ & $F$ & $F_0^*$ & $F_1^*$ & $F_2^*$ & $F$ \\
        \hline
        \multirow{2}{1em}{$\theta_1$} & $0.979$ & $0.979$ & $1.000$ & $0.962$ & $0.981$ & $0.981$ & $1.000$ & $0.992$ & $0.661$ & $0.782$ & $0.782$ & $0.960$\\
        \cline{2-13}
         & \multicolumn{4}{|c|}{$r^* \geq 1$ (WN)} & \multicolumn{4}{|c|}{$r^* \geq 2$} & \multicolumn{4}{|c|}{$r^* \geq 3$} \\
        \hline
        \multirow{2}{1em}{$\theta_2$} & $0.988$ & $0.988$ & $1.000$ & $0.854$ & $0.972$ & $0.972$ & $1.000$ & $0.998$ & $0.692$ & $1.000$ & $1.000$ & $0.995$\\
        \cline{2-13}
         & \multicolumn{4}{|c|}{$r^* \geq 1$ (WN)} & \multicolumn{4}{|c|}{$r^* \geq 2$} & \multicolumn{4}{|c|}{$r^* \geq 1$} \\
        \hline
        \multirow{2}{1em}{$\theta_3$} & $0.957$ & $0.957$ & $1.000$ & $0.902$ & $0.963$ & $0.963$ & $1.000$ & $0.976$ & $0.706$ & $0.967$ & $0.967$ & $0.993$\\
        \cline{2-13}
         & \multicolumn{4}{|c|}{$r^* \geq 1$ (WN)} & \multicolumn{4}{|c|}{$r^* \geq 2$} & \multicolumn{4}{|c|}{$r^* \geq 3$} \\
        \hline
    \end{tabular}
    \caption{Stellar rank witnesses for the different states described in fig \ref{fig:squeezed_input}, obtained from a squeezed vacuum input. Here $F = F(\rho^\text{core}, \Psi^\text{core})$ and $F_k^* = 1 - \bigl(R_k^*(\Psi^\text{core})\bigr)^2$, where $R_k^*(\Psi^\text{core})$ is the k-robustness of $\ket{\Psi^\text{core}}$ \cite{ChabaudCertification2021}. The bounds on the stellar rank $r^*$ are obtained according to relation \ref{eq:stellar_rank_bound} or thanks to Wigner negativity (WN). Values of $F_k^*$ are rounded up while values of $F$ are rounded down.
    }
    \label{tab:rank_witnesses}
\end{table*}

\section{Core states and stellar rank} \label{supp:stellar_rank}

Any pure state of light can be written in the following way:
\begin{equation}
    \ket{\Psi} = U_G \sum a_n \ket{n}
\end{equation}
Where $U_G$ is a product of Gaussian operators (displacement or squeezing). To find the core state, we are looking for an operator $U_G^*$ that minimizes the integer r such that
\begin{equation}
    \ket{\Psi} = U_G^* \sum_{n=0}^{r^*} b_{n} \ket{n} =  U_G^* \ket{\psi^\text{core}}
\end{equation}
Here $r^*$ corresponds to the stellar rank of the quantum state $\ket{\Psi}$. To extend this definition to mixed states, one needs to consider all possible decompositions of the mixed state into pure states, making it much harder to compute \cite{ChabaudStellar2020}. Here, we use fidelity witnesses to bound the stellar rank of the obtained states, as presented in \cite{ChabaudCertification2021}. 

Let's consider a given output state $\rho$. We first estimate the core state by minimizing the following loss function: $f(\alpha, z) = \sum_n (n+1)^p \sqrt{\bra{n}\rho'(\alpha, z) \ket{n} + \epsilon}$ where $\rho'(\alpha, z) = \D(\alpha)\hat{S}(z) \rho \D(-\alpha)\hat{S}(-z)$ and $\epsilon = 10^-5$. This empirical formula is inspired by weighted $\mathcal{L}_1$ norms sometimes used in data analysis \cite{candes2008enhancing}. It penalizes contributions of large Fock states through the weights $(n + 1)^p$, where $p$ is an arbitrary integer that we can adjust. It also disproportionally penalizes contributions of small amplitude thanks to the square root. We add $\epsilon$ to stop gradients from diverging.
The density matrix $\rho'$ that minimizes $f$ should therefore be a scarce matrix that involves the smallest possible Fock states. To find this minimum, we first compute the loss function on a coarse-grained array of values of $\alpha$ and $z$ to estimate the position of the global minimum. We then perform a gradient descent on parameters $\alpha$ and $z$ to obtain a more precise result. 

This core state estimate is used in Fig. \ref{fig:coherent_input} to characterize the output state. 

We ``purify'' the core state $\rho^\mathrm{core}$ by taking the pure state $\ket{\Psi^\text{core}}$ that maximizes $F(\rho^\mathrm{core}, \Psi^\mathrm{core})$. $\ket{\Psi^\text{core}}$ is the eigenvector of the density matrix with the largest associated eigenvalue \cite{JozsaFidelity}. We then compute the k-robustness of $\Psi^\text{core}$ for $k=0$, $1$ and $2$. It is defined as \cite{ChabaudCertification2021} $$R_k^*(\Psi) \equiv \underset{{r^*(\Phi)<k}}{\text{inf}} \sqrt{1 - F(\Phi, \Psi}) $$
Where $F(\Phi, \Psi)$ is the fidelity between $\ket{\Phi}$ and $\ket{\Psi}$.
Thus, to compute the k-robustness of $\Psi^\text{core}$, we maximize the fidelity between $\ket{\Phi}(\alpha, z, c_1,...,c_k) = \hat{S}(z)\D(\alpha)(c_0 \ket{0} + c_1 \ket{1} + ... + c_k \ket{k})$ and $\ket{\Psi^\text{core}}$. As before, we first find a rough estimate of the global minimum and we optimize it by performing a gradient descent on parameters $(\alpha, z, c_1,...,c_k)$. 

We then compare the quantities $F(\rho^\text{core}, \Psi^\text{core})$ and $F_k^*(\Psi^\text{core}) = 1 - \bigl(R_k^*(\Psi^\text{core})\bigr)^2$ to bound the rank of $\rho$. Indeed: 
\begin{equation}\label{eq:stellar_rank_bound}
    F(\rho^\text{core}, \Psi^\text{core}) > F_k^*(\Psi^\text{core}) \implies r^*(\rho) > k
\end{equation}
If $F(\rho^\text{core}, \Psi^\text{core}) \leq F_0^*(\Psi^\text{core})$, the presence of Wigner negativity is enough to claim $r^*(\rho) \geq 1$.

Table \ref{tab:rank_witnesses} gathers all the results obtained with this method on the states described in Fig. \ref{fig:squeezed_input}. For $\theta_1$, $\theta_2$ and $\theta_3$, the full output state $\rho_{1, \theta}$ (in mode $v_1^\theta$) is not pure enough to guarantee a stellar rank $r^*>1$ with this method. However, we know that all states studied here have a stellar rank $r^* \geq 1$, as they all display Wigner negativity. 

$\rho_{1, \theta}^\text{even}$ has a stellar rank $r^* \geq 2$ for the three different bases. For $\theta_1$ and $\theta_3$, $\rho_{1, \theta}^\text{odd}$ has a stellar rank $r^* \geq 3$. Conversely, $\rho_{1, \theta_2}^\text{odd}$ is a nearly pure squeezed single photon, which is consistent with $r^* \geq 1$. 

We have not proven that our strategy to find the core state is optimal, but it is good enough to find a pure state that has a high fidelity with a given output state $\rho$, and thus establish bounds on its stellar rank.

\section{Quantum Non Gaussian Coherences}
In Sec.~\ref{sec:coherent_input}, photon-number coherences are used to characterize the generated core states, providing a complementary perspective to stellar-rank based witnesses. A similar analysis can be done on states from a squeezed-vacuum input.

Here, instead of searching for the Gaussian transformation that maps the output state onto a minimal core state, we look for the transformation that maximizes selected photon-number coherences. Following the approach developed in Sec.~\ref{sec:squeezed_input}, we start from $\rho_{1, \theta}$ and decompose it into its even and odd components $\rho_{1,\theta}^\mathrm{even}$ and $\rho_{1,\theta}^\mathrm{odd}$, where $\theta\in [-\pi/2,\pi/2]$ parametrizes the output-mode basis rotation. For each value of $\theta$ and for each of the three corresponding density matrices, we optimize on the Gaussian operation $U_G$ to maximize the relevant coherence terms. For $\rho = \rho_{1, \theta}$ and $\rho = \rho_{1,\theta}^\mathrm{even}$, we maximize the term $\mathcal{C}_{0,2} = 2\abs{\bra{0} U_G \rho U_G^\dag \ket{2}}$. For $\rho = \rho_{1,\theta}^\mathrm{odd}$, we maximize $\mathcal{C}_{1,3} = 2\abs{\bra{1} U_G \rho U_G^\dag \ket{3}}$. For both investigated quantities, the quantum non-Gaussian coherence threshold is $0.86$ \cite{LachmanCoherences, TejaQNG2025}.

\begin{figure}[t]
    \centering
    \includegraphics[width=0.6\linewidth]{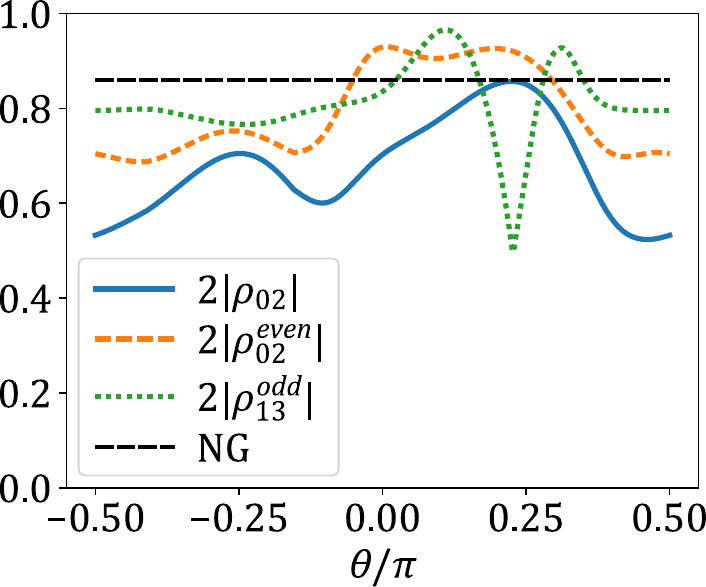}
    \caption{
    Maximum quantum coherences of $\rho_{1, \theta}$, $\rho_{1,\theta}^\mathrm{even}$ and $\rho_{1,\theta}^\mathrm{odd}$ as a function of $\theta$. For each value of $\theta$ and each density matrix $\rho_{1, \theta}$, $\rho_{1,\theta}^\mathrm{even}$ and $\rho_{1,\theta}^\mathrm{odd}$, a Gaussian transformation $U$ is chosen to maximize the coherence measurements $\mathcal{C}_{0,2} =  2\abs{\bra{0} U \rho U^\dag \ket{2}}$ or $\mathcal{C}_{1,3} =  2\abs{\bra{1} U \rho U^\dag \ket{3}}$. The dashed line at $0.86$ corresponds to the quantum non-Gaussian coherence threshold. The threshold is surpassed for $\rho_{1,\theta}^\mathrm{even}$ and $\rho_{1,\theta}^\mathrm{odd}$, but not for $\rho_{1,\theta}$. 
    }
    \label{fig:coherences}
\end{figure}

In Fig. \ref{fig:coherences}, we plot the maximal coherences for $\rho_{1, \theta}$, $\rho_{1,\theta}^\mathrm{even}$ and $\rho_{1,\theta}^\mathrm{odd}$ as a function of $\theta$. For $\rho_{1, \theta}$, $\mathcal{C}_{0,2}$ is maximal for $\theta = 0.224 \pi$ and reaches $\underset{\theta}{\text{max }}\mathcal{C}_{0,2} = 0.857$, just below the QNG coherence threshold. For $\rho_{1,\theta}^\mathrm{even}$, the threshold is exceeded on a broad range of $\theta$, and we have $\underset{\theta}{\text{max }}\mathcal{C}_{0,2} = 0.930$, reached for $\theta = 0.006 \pi$. Finally, for $\rho_{1,\theta}^\mathrm{odd}$, the threshold is also exceeded, with $\underset{\theta}{\text{max }}\mathcal{C}_{1,3} = 0.966$, reached at $\theta = 0.110 \pi$.

As can be expected from the structure of the density matrices $\rho_{1, \theta}$, $\rho_{1,\theta}^\mathrm{even}$ and $\rho_{1,\theta}^\mathrm{odd}$, centered in phase-space, the best Gaussian transformation $U_G$ is always pure squeezing.

\section{Numerical simulations} \label{supp:numerical_sims}
The simulations are implemented using the QuTiP library on Python \cite{johansson2012qutip, qutip5}. In order to compute the evolution of the system, we need to introduce up to four Hilbert spaces corresponding to the following objects : input cavity, TLS, output cavity 1, output cavity 2. We introduce a maximum number of photons $N-1$ in each cavity. The total Hilbert space is of dimension $d = 2 N^3$. An operator in this space has $4 N^6$ coefficients. For squeezed vacuum inputs, we set $N = 13$, leading to a Hilbert space of dimension $d = 4394$. For coherent inputs, we can eliminate the input cavity by adding a classical driving term in the Hamiltonian \cite{kiilerich_quantum_2020}. We set $N=15$ for coherent inputs, leading to a Hilbert space of dimension $d = 2N^2 = 450$. 

\section{Definitions} \label{supp:definitions}
\noindent\textbf{Wigner negativity volume:}

In this work, we define the Wigner negativity volume of a quantum state as the volume of the negative part of its Wigner function \cite{Kenfack_2004}.
\begin{equation} \label{eq:negativity_def}
    \mathcal{N} = \frac{1}{2} \int \dd^2 \alpha \left(\abs{\mathcal{W}(\alpha)} - \mathcal{W}(\alpha)\right)
\end{equation}

\noindent\textbf{Fidelity:}

For two quantum states $\rho$ and $\sigma$, the fidelity is defined as \cite{JozsaFidelity} $F(\rho, \sigma) = \Tr(\sqrt{\sqrt{\rho}\sigma\sqrt{\rho}})^2$.
It simplifies to $F(\rho, \psi) = \bra{\psi} \rho \ket{\psi}$ if $\sigma = \ket{\psi}\bra{\psi}$. It becomes $F(\psi, \phi) = |\langle \psi | \phi \rangle|^2$ for two pure states $\ket{\psi}$ and $\ket{\phi}$.

\bibliographystyle{apsrev4-2-short}
\bibliography{references}

@article{larsen_deterministic_2019,
	title = {Deterministic generation of a two-dimensional cluster state},
	volume = {366},
	issn = {0036-8075, 1095-9203},
	url = {http://arxiv.org/abs/1906.08709},
	doi = {10.1126/science.aay4354},
	number = {6463},
	urldate = {2026-06-26},
	journal = {Science},
	author = {Larsen, Mikkel V. and Guo, Xueshi and Breum, Casper R. and Neergaard-Nielsen, Jonas S. and Andersen, Ulrik L.},
	month = oct,
	year = {2019},
	note = {arXiv:1906.08709 [quant-ph]},
	pages = {369--372},
}

@article{larsen_GKP_2025,
	title = {Integrated photonic source of {Gottesman}–{Kitaev}–{Preskill} qubits},
	volume = {642},
	copyright = {2025 The Author(s)},
	issn = {1476-4687},
	url = {https://www.nature.com/articles/s41586-025-09044-5},
	doi = {10.1038/s41586-025-09044-5},
	language = {english},
	number = {8068},
	urldate = {2026-05-05},
	journal = {Nature},
	publisher = {Nature Publishing Group},
	author = {Larsen, M. V. and Bourassa, J. E. and Kocsis, S. and Tasker, J. F. and Chadwick, R. S. and González-Arciniegas, C. and Hastrup, J. and Lopetegui-González, C. E. and Miatto, F. M. and Motamedi, A. and Noro, R. and Roeland, G. and Baby, R. and Chen, H. and Contu, P. and Di Luch, I. and Drago, C. and Giesbrecht, M. and Grainge, T. and Krasnokutska, I. and Menotti, M. and Morrison, B. and Puviraj, C. and Rezaei Shad, K. and Hussain, B. and McMahon, J. and Ortmann, J. E. and Collins, M. J. and Ma, C. and Phillips, D. S. and Seymour, M. and Tang, Q. Y. and Yang, B. and Vernon, Z. and Alexander, R. N. and Mahler, D. H.},
	month = jun,
	year = {2025},
	pages = {587--591}
}

@article{GKP2001,
  title = {Encoding a qubit in an oscillator},
  author = {Gottesman, Daniel and Kitaev, Alexei and Preskill, John},
  journal = {Phys. Rev. A},
  volume = {64},
  issue = {1},
  pages = {012310},
  numpages = {21},
  year = {2001},
  month = {Jun},
  publisher = {American Physical Society},
  doi = {10.1103/PhysRevA.64.012310},
  url = {https://link.aps.org/doi/10.1103/PhysRevA.64.012310}
}

@article{BartlettGaussianStateSimulation,
  title = {Efficient Classical Simulation of Continuous Variable Quantum Information Processes},
  author = {Bartlett, Stephen D. and Sanders, Barry C. and Braunstein, Samuel L. and Nemoto, Kae},
  journal = {Phys. Rev. Lett.},
  volume = {88},
  issue = {9},
  pages = {097904},
  numpages = {4},
  year = {2002},
  month = {Feb},
  publisher = {American Physical Society},
  doi = {10.1103/PhysRevLett.88.097904},
  url = {https://link.aps.org/doi/10.1103/PhysRevLett.88.097904}
}

@misc{gottesman1998heisenbergrepresentationquantumcomputers,
      title={The Heisenberg Representation of Quantum Computers}, 
      author={Daniel Gottesman},
      year={1998},
      eprint={quant-ph/9807006},
      archivePrefix={arXiv},
      primaryClass={quant-ph},
      url={https://arxiv.org/abs/quant-ph/9807006}, 
}

@book{haroche_exploring_2006,
	title = {Exploring the {Quantum}: {Atoms}, {Cavities}, and {Photons}},
	isbn = {978-0-19-850914-1},
	url = {https://doi.org/10.1093/acprof:oso/9780198509141.001.0001},
	urldate = {2024-07-23},
	publisher = {Oxford University Press},
	author = {Haroche, Serge and Raimond, Jean-Michel},
	month = aug,
	year = {2006},
	doi = {10.1093/acprof:oso/9780198509141.001.0001},
}

@article{Honer_2011,
  title = {Artificial Atoms Can Do More Than Atoms: Deterministic Single Photon Subtraction from Arbitrary Light Fields},
  author = {Honer, Jens and L\"ow, R. and Weimer, Hendrik and Pfau, Tilman and B\"uchler, Hans Peter},
  journal = {Phys. Rev. Lett.},
  volume = {107},
  issue = {9},
  pages = {093601},
  numpages = {5},
  year = {2011},
  month = {Aug},
  publisher = {American Physical Society},
  doi = {10.1103/PhysRevLett.107.093601},
  url = {https://link.aps.org/doi/10.1103/PhysRevLett.107.093601}
}

@article{Hacker2019,
  title = {Deterministic creation of entangled atom–light Schr\"{o}dinger-cat states},
  volume = {13},
  ISSN = {1749-4893},
  url = {http://dx.doi.org/10.1038/s41566-018-0339-5},
  DOI = {10.1038/s41566-018-0339-5},
  number = {2},
  journal = {Nature Photonics},
  publisher = {Springer Science and Business Media LLC},
  author = {Hacker,  Bastian and Welte,  Stephan and Daiss,  Severin and Shaukat,  Armin and Ritter,  Stephan and Li,  Lin and Rempe,  Gerhard},
  year = {2019},
  month = Jan,
  pages = {110–115}
}

@article{magro_deterministic_2023,
	title = {Deterministic freely propagating photonic qubits with negative {Wigner} functions},
	volume = {17},
	copyright = {2023 The Author(s), under exclusive licence to Springer Nature Limited},
	issn = {1749-4893},
	url = {https://www.nature.com/articles/s41566-023-01196-y},
	doi = {10.1038/s41566-023-01196-y},
	language = {english},
	number = {8},
	urldate = {2026-05-05},
	journal = {Nature Photonics},
	publisher = {Nature Publishing Group},
	author = {Magro, Valentin and Vaneecloo, Julien and Garcia, Sébastien and Ourjoumtsev, Alexei},
	month = aug,
	year = {2023},
	pages = {688--693},
}

@article{Gonzalez-Tudela_2015_Det,
  title = {Deterministic Generation of Arbitrary Photonic States Assisted by Dissipation},
  author = {Gonz\'alez-Tudela, A. and Paulisch, V. and Chang, D. E. and Kimble, H. J. and Cirac, J. I.},
  journal = {Phys. Rev. Lett.},
  volume = {115},
  issue = {16},
  pages = {163603},
  numpages = {6},
  year = {2015},
  month = {Oct},
  publisher = {American Physical Society},
  doi = {10.1103/PhysRevLett.115.163603},
  url = {https://link.aps.org/doi/10.1103/PhysRevLett.115.163603}
}

@article{Paulisch_2018,
  title = {Generation of single- and two-mode multiphoton states in waveguide QED},
  author = {Paulisch, V. and Kimble, H. J. and Cirac, J. I. and Gonz\'alez-Tudela, A.},
  journal = {Phys. Rev. A},
  volume = {97},
  issue = {5},
  pages = {053831},
  numpages = {15},
  year = {2018},
  month = {May},
  publisher = {American Physical Society},
  doi = {10.1103/PhysRevA.97.053831},
  url = {https://link.aps.org/doi/10.1103/PhysRevA.97.053831}
}

@article{lund_subtraction_2024,
	title = {Subtraction and {Addition} of {Propagating} {Photons} by {Two}-{Level} {Emitters}},
	volume = {133},
	url = {https://link.aps.org/doi/10.1103/PhysRevLett.133.103601},
	doi = {10.1103/PhysRevLett.133.103601},
	number = {10},
	urldate = {2025-05-06},
	journal = {Physical Review Letters},
	publisher = {American Physical Society},
	author = {Lund, Mads M. and Yang, Fan and Christiansen, Victor Rueskov and Kornovan, Danil and Mølmer, Klaus},
	month = sep,
	year = {2024},
	pages = {103601},
}

@article{pasharavesh_generation_2024,
	title = {Generation of non-{Gaussian} states of light using deterministic photon subtraction},
	volume = {26},
	issn = {1367-2630},
	url = {https://iopscience.iop.org/article/10.1088/1367-2630/ad95b3},
	doi = {10.1088/1367-2630/ad95b3},
	language = {english},
	number = {11},
	urldate = {2025-10-14},
	journal = {New Journal of Physics},
	author = {Pasharavesh, Abdolreza and Bajcsy, Michal},
	month = nov,
	year = {2024},
	pages = {113022},
}

@misc{luo2025dynamicstimulatedemissiondeterministic,
      title={Dynamic stimulated emission for deterministic addition and subtraction of propagating photons}, 
      author={Haoyuan Luo and Parth S. Shah and Frank Yang and Mohammad Mirhosseini and Sahand Mahmoodian},
      year={2025},
      eprint={2512.09711},
      archivePrefix={arXiv},
      primaryClass={quant-ph},
      url={https://arxiv.org/abs/2512.09711}, 
}

@misc{Bouchereau2026,
      title={Temporal modes of quantum states of light scattered by a two-level system}, 
      author={Yann Bouchereau and Lucas Weitzel and Valerian Thiel and Valentina Parigi and Mattia Walschaers and Nicolas Treps},
      year={2026},
      eprint={2606.29974},
      archivePrefix={arXiv},
      primaryClass={quant-ph},
      url={https://arxiv.org/abs/2606.29974}, 
}

@article{kleinbeck_creation_2023,
	title = {Creation of nonclassical states of light in a chiral waveguide},
	volume = {107},
	url = {https://link.aps.org/doi/10.1103/PhysRevA.107.013717},
	doi = {10.1103/PhysRevA.107.013717},
	number = {1},
	urldate = {2025-04-07},
	journal = {Physical Review A},
	publisher = {American Physical Society},
	author = {Kleinbeck, Kevin and Busche, Hannes and Stiesdal, Nina and Hofferberth, Sebastian and Mølmer, Klaus and Büchler, Hans Peter},
	month = jan,
	year = {2023},
	pages = {013717}}

@article{quijandria_steady-state_2018,
	title = {Steady-{State} {Generation} of {Wigner}-{Negative} {States} in {One}-{Dimensional} {Resonance} {Fluorescence}},
	volume = {121},
	url = {https://link.aps.org/doi/10.1103/PhysRevLett.121.263603},
	doi = {10.1103/PhysRevLett.121.263603},
	number = {26},
	urldate = {2025-06-11},
	journal = {Physical Review Letters},
	publisher = {American Physical Society},
	author = {Quijandría, Fernando and Strandberg, Ingrid and Johansson, Göran},
	month = dec,
	year = {2018},
	pages = {263603},
}

@article{leonhardt_wigner-negative_2025,
	title = {Wigner-negative states in the steady-state emission of a two-level system driven by squeezed light},
	volume = {112},
	issn = {2469-9926, 2469-9934},
	url = {http://arxiv.org/abs/2408.01698},
	doi = {10.1103/1kbt-kqjd},
	number = {1},
	urldate = {2026-02-03},
	journal = {Physical Review A},
	author = {Leonhardt, Miriam J. and Parkins, Scott},
	month = jul,
	year = {2025},
	note = {arXiv:2408.01698 [quant-ph]},
	pages = {013715}}

@article{Luo_2026,
  title = {Efficient generation of optical cat states using squeezed few-photon superposition states},
  author = {Luo, Haoyuan and Mahmoodian, Sahand},
  journal = {Phys. Rev. Appl.},
  volume = {25},
  issue = {2},
  pages = {024046},
  numpages = {20},
  year = {2026},
  month = {Feb},
  publisher = {American Physical Society},
  doi = {10.1103/t96h-488y},
  url = {https://link.aps.org/doi/10.1103/t96h-488y}
}

@article{lodahl_chiral_2017,
	title = {Chiral {Quantum} {Optics}},
	volume = {541},
	issn = {0028-0836, 1476-4687},
	url = {http://arxiv.org/abs/1608.00446},
	doi = {10.1038/nature21037},
	language = {english},
	number = {7638},
	urldate = {2025-12-12},
	journal = {Nature},
	author = {Lodahl, Peter and Mahmoodian, Sahand and Stobbe, Søren and Schneeweiss, Philipp and Volz, Jürgen and Rauschenbeutel, Arno and Pichler, Hannes and Zoller, Peter},
	month = jan,
	year = {2017},
	note = {arXiv:1608.00446 [quant-ph]},
	pages = {473--480}}

@book{Scully_Zubairy_1997, place={Cambridge}, title={Quantum Optics}, publisher={Cambridge University Press}, author={Scully, Marlan O. and Zubairy, M. Suhail}, year={1997}}

@misc{prasad_closing_2024,
	title = {Closing {Optical} {Bloch} {Equations} in waveguide {QED}: {Dynamics}, {Energetics}},
    shorttitle = {Closing {Optical} {Bloch} {Equations} in waveguide {QED}},
	url = {https://arxiv.org/abs/2404.09648v1},
	language = {english},
	urldate = {2025-06-20},
	journal = {arXiv.org},
	author = {Prasad, Samyak Pratyush and Maffei, Maria and Camati, Patrice A. and Elouard, Cyril and Auffèves, Alexia},
	month = apr,
	year = {2024},
}

@article{gardiner_input_1985,
	title = {Input and output in damped quantum systems: {Quantum} stochastic differential equations and the master equation},
	volume = {31},
	copyright = {http://link.aps.org/licenses/aps-default-license},
	issn = {0556-2791},
	shorttitle = {Input and output in damped quantum systems},
	url = {https://link.aps.org/doi/10.1103/PhysRevA.31.3761},
	doi = {10.1103/PhysRevA.31.3761},
	language = {english},
	number = {6},
	urldate = {2024-06-12},
	journal = {Physical Review A},
	author = {Gardiner, C. W. and Collett, M. J.},
	month = jun,
	year = {1985},
	pages = {3761--3774},
}

@article{kiilerich_input-output_2019,
	title = {Input-{Output} {Theory} with {Quantum} {Pulses}},
	volume = {123},
	url = {https://link.aps.org/doi/10.1103/PhysRevLett.123.123604},
	doi = {10.1103/PhysRevLett.123.123604},
	number = {12},
	urldate = {2025-04-23},
	journal = {Physical Review Letters},
	publisher = {American Physical Society},
	author = {Kiilerich, Alexander Holm and Mølmer, Klaus},
	month = sep,
	year = {2019},
	pages = {123604},
}

@article{kiilerich_quantum_2020,
	title = {Quantum interactions with pulses of radiation},
	volume = {102},
	url = {https://link.aps.org/doi/10.1103/PhysRevA.102.023717},
	doi = {10.1103/PhysRevA.102.023717},
	number = {2},
	urldate = {2025-04-07},
	journal = {Physical Review A},
	publisher = {American Physical Society},
	author = {Kiilerich, Alexander Holm and Mølmer, Klaus},
	month = aug,
	year = {2020},
	pages = {023717},
}

@article{Brecht2014,
  title = {Demonstration of coherent time-frequency Schmidt mode selection using dispersion-engineered frequency conversion},
  author = {Brecht, Benjamin and Eckstein, Andreas and Ricken, Raimund and Quiring, Viktor and Suche, Hubertus and Sansoni, Linda and Silberhorn, Christine},
  journal = {Phys. Rev. A},
  volume = {90},
  issue = {3},
  pages = {030302(R)},
  numpages = {5},
  year = {2014},
  month = {Sep},
  publisher = {American Physical Society},
  doi = {10.1103/PhysRevA.90.030302},
  url = {https://link.aps.org/doi/10.1103/PhysRevA.90.030302}
}

@misc{Lim2026,
      title={Optimizing Wigner Negativity in Scattering Processes Using Energetic Cost Functions}, 
      author={Kian Hwee Lim and Kiarn T. Laverick and Sahil Sardar Jafar and Samyak P. Prasad and Maria Maffei and Alexia Auffèves},
      year={2026},
      eprint={2606.15101},
      archivePrefix={arXiv},
      primaryClass={quant-ph},
      url={https://arxiv.org/abs/2606.15101}, 
}

@article{Menzies2009,
  title = {Gaussian-optimized preparation of non-Gaussian pure states},
  author = {Menzies, David and Filip, Radim},
  journal = {Phys. Rev. A},
  volume = {79},
  issue = {1},
  pages = {012313},
  numpages = {7},
  year = {2009},
  month = {Jan},
  publisher = {American Physical Society},
  doi = {10.1103/PhysRevA.79.012313},
  url = {https://link.aps.org/doi/10.1103/PhysRevA.79.012313}
}

@article{Lachman2019,
  title = {Faithful Hierarchy of Genuine $n$-Photon Quantum Non-Gaussian Light},
  author = {Lachman, Luk\'a\v{s} and Straka, Ivo and Hlou\v{s}ek, Josef and Je\ifmmode \check{z}\else \v{z}\fi{}ek, Miroslav and Filip, Radim},
  journal = {Phys. Rev. Lett.},
  volume = {123},
  issue = {4},
  pages = {043601},
  numpages = {6},
  year = {2019},
  month = {Jul},
  publisher = {American Physical Society},
  doi = {10.1103/PhysRevLett.123.043601},
  url = {https://link.aps.org/doi/10.1103/PhysRevLett.123.043601}
}

@article{Fiurasek2022,
author = {Jarom\'{i}r Fiur\'{a}\v{s}ek},
journal = {Opt. Express},
number = {17},
pages = {30630--30639},
publisher = {Optica Publishing Group},
title = {Efficient construction of witnesses of the stellar rank of nonclassical states of light},
volume = {30},
month = {Aug},
year = {2022},
url = {https://opg.optica.org/oe/abstract.cfm?URI=oe-30-17-30630},
doi = {10.1364/OE.466175},
}

@article{Chabaud2023,
  title = {Resources for Bosonic Quantum Computational Advantage},
  author = {Chabaud, Ulysse and Walschaers, Mattia},
  journal = {Phys. Rev. Lett.},
  volume = {130},
  issue = {9},
  pages = {090602},
  numpages = {7},
  year = {2023},
  month = {Mar},
  publisher = {American Physical Society},
  doi = {10.1103/PhysRevLett.130.090602},
  url = {https://link.aps.org/doi/10.1103/PhysRevLett.130.090602}
}

@article{Ast_2013,
   title={High-bandwidth squeezed light at 1550 nm from a compact monolithic PPKTP cavity},
   volume={21},
   ISSN={1094-4087},
   url={http://dx.doi.org/10.1364/OE.21.013572},
   DOI={10.1364/oe.21.013572},
   number={11},
   journal={Optics Express},
   publisher={Optica Publishing Group},
   author={Ast, Stefan and Mehmet, Moritz and Schnabel, Roman},
   year={2013},
   month=May, pages={13572} }

@misc{terrasson2026,
      title={Bright Pulsed Squeezed Light for Quantum-Enhanced Precision Microscopy}, 
      author={Alex Terrasson and Lars Madsen and Joel Grim and Warwick Bowen},
      year={2026},
      eprint={2601.15565},
      archivePrefix={arXiv},
      primaryClass={quant-ph},
      url={https://arxiv.org/abs/2601.15565}, 
}

@article{Eckstein:11,
author = {Andreas Eckstein and Benjamin Brecht and Christine Silberhorn},
journal = {Opt. Express},
number = {15},
pages = {13770--13778},
publisher = {Optica Publishing Group},
title = {A quantum pulse gate based on spectrally engineered sum frequency generation},
volume = {19},
month = {Jul},
year = {2011},
url = {https://opg.optica.org/oe/abstract.cfm?URI=oe-19-15-13770},
doi = {10.1364/OE.19.013770},
}

@article{SerinoMQPG2023,
  title = {Realization of a Multi-Output Quantum Pulse Gate for Decoding High-Dimensional Temporal Modes of Single-Photon States},
  author = {Serino, Laura and Gil-Lopez, Jano and Stefszky, Michael and Ricken, Raimund and Eigner, Christof and Brecht, Benjamin and Silberhorn, Christine},
  journal = {PRX Quantum},
  volume = {4},
  issue = {2},
  pages = {020306},
  numpages = {12},
  year = {2023},
  month = {Apr},
  publisher = {American Physical Society},
  doi = {10.1103/PRXQuantum.4.020306},
  url = {https://link.aps.org/doi/10.1103/PRXQuantum.4.020306}
}

@article{Christiansen2026,
  title = {Interactions in quantum networks with pulse propagation delays},
  author = {Christiansen, Victor Rueskov and M\o{}lmer, Klaus},
  journal = {Phys. Rev. A},
  volume = {113},
  issue = {1},
  pages = {013730},
  numpages = {8},
  year = {2026},
  month = {Jan},
  publisher = {American Physical Society},
  doi = {10.1103/3f3w-jmj8},
  url = {https://link.aps.org/doi/10.1103/3f3w-jmj8}
}

@article{Konno_2024,
author = {Shunya Konno  and Warit Asavanant  and Fumiya Hanamura  and Hironari Nagayoshi  and Kosuke Fukui  and Atsushi Sakaguchi  and Ryuhoh Ide  and Fumihiro China  and Masahiro Yabuno  and Shigehito Miki  and Hirotaka Terai  and Kan Takase  and Mamoru Endo  and Petr Marek  and Radim Filip  and Peter van Loock  and Akira Furusawa },
title = {Logical states for fault-tolerant quantum computation with propagating light},
journal = {Science},
volume = {383},
number = {6680},
pages = {289-293},
year = {2024},
doi = {10.1126/science.adk7560},
URL = {https://www.science.org/doi/abs/10.1126/science.adk7560},
eprint = {https://www.science.org/doi/pdf/10.1126/science.adk7560}}

@article{Arcari2014,
  title = {Near-Unity Coupling Efficiency of a Quantum Emitter to a Photonic Crystal Waveguide},
  author = {Arcari, M. and S\"ollner, I. and Javadi, A. and Lindskov Hansen, S. and Mahmoodian, S. and Liu, J. and Thyrrestrup, H. and Lee, E. H. and Song, J. D. and Stobbe, S. and Lodahl, P.},
  journal = {Phys. Rev. Lett.},
  volume = {113},
  issue = {9},
  pages = {093603},
  numpages = {5},
  year = {2014},
  month = {Aug},
  publisher = {American Physical Society},
  doi = {10.1103/PhysRevLett.113.093603},
  url = {https://link.aps.org/doi/10.1103/PhysRevLett.113.093603}
}

@article{lahadInducedCavitiesPhotonic2017,
	title = {Induced {Cavities} for {Photonic} {Quantum} {Gates}},
	volume = {119},
	copyright = {https://link.aps.org/licenses/aps-default-license},
	issn = {0031-9007, 1079-7114},
	url = {https://link.aps.org/doi/10.1103/PhysRevLett.119.113601},
	doi = {10.1103/PhysRevLett.119.113601},
	language = {english},
	number = {11},
	urldate = {2026-01-05},
	journal = {Physical Review Letters},
	author = {Lahad, Ohr and Firstenberg, Ofer},
	month = sep,
	year = {2017},
	pages = {113601}}

@article{Yang_2025,
  title   = {Entanglement of photonic modes from a continuously driven two-level system},
  author  = {Yang, Jiaying and Strandberg, Ingrid and Vivas-Via{\~n}a, Alejandro and Gaikwad, Akshay and Castillo-Moreno, Claudia and Kockum, Anton Frisk and Ullah, Muhammad Asad and S{\'a}nchez Mu{\~n}oz, Carlos and Eriksson, Axel Martin and Gasparinetti, Simone},
  journal = {npj Quantum Information},
  volume  = {11},
  pages   = {69},
  year    = {2025},
  doi     = {10.1038/s41534-025-00995-1}
}

@article{Kiryu_2026,
  title   = {Linear-Optical Generation of Hybrid {GKP} Entanglement from Small-Amplitude Cat States},
  author  = {Kiryu, Shohei and Chin, Yohji and Takeoka, Masahiro and Fukui, Kosuke},
  journal = {arXiv preprint arXiv:2603.19870},
  year    = {2026},
  doi     = {10.48550/arXiv.2603.19870},
  url     = {https://arxiv.org/abs/2603.19870}
}

@article{Ding2025,
  title = {Photon-Number-Resolving Single-Photon Detector with a System Detection Efficiency of 98% and Photon-Number Resolution of 32},
  volume = {12},
  ISSN = {2330-4022},
  url = {http://dx.doi.org/10.1021/acsphotonics.5c00508},
  DOI = {10.1021/acsphotonics.5c00508},
  number = {9},
  journal = {ACS Photonics},
  publisher = {American Chemical Society (ACS)},
  author = {Ding,  Chaomeng and Zhang,  Xingyu and Xiong,  Jiamin and Xiao,  You and Zhang,  Tianzhu and Huang,  Jia and Xu,  Hongxin and Liu,  Xiaoyu and You,  Lixing and Wang,  Zhen and Li,  Hao},
  year = {2025},
  month = {July},
  pages = {4924–4931}
}

@article{Eaton_2019,
  title   = {Non-Gaussian and Gottesman-Kitaev-Preskill State Preparation by Photon Catalysis},
  author  = {Eaton, Miller and Nehra, Rajveer and Pfister, Olivier},
  journal = {New Journal of Physics},
  volume  = {21},
  number  = {11},
  pages   = {113034},
  year    = {2019},
  doi     = {10.1088/1367-2630/ab5330}
}

@article{Winnel_2024,
  title = {Deterministic Preparation of Optical Squeezed Cat and Gottesman-Kitaev-Preskill States},
  author = {Winnel, Matthew S. and Guanzon, Joshua J. and Singh, Deepesh and Ralph, Timothy C.},
  journal = {Phys. Rev. Lett.},
  volume = {132},
  issue = {23},
  pages = {230602},
  numpages = {6},
  year = {2024},
  month = {Jun},
  publisher = {American Physical Society},
  doi = {10.1103/PhysRevLett.132.230602},
  url = {https://link.aps.org/doi/10.1103/PhysRevLett.132.230602}
}

@article{ChabaudCertification2021,
  title = {Certification of Non-Gaussian States with Operational Measurements},
  author = {Chabaud, Ulysse and Roeland, Gana\"el and Walschaers, Mattia and Grosshans, Fr\'ed\'eric and Parigi, Valentina and Markham, Damian and Treps, Nicolas},
  journal = {PRX Quantum},
  volume = {2},
  issue = {2},
  pages = {020333},
  numpages = {19},
  year = {2021},
  month = {Jun},
  publisher = {American Physical Society},
  doi = {10.1103/PRXQuantum.2.020333},
  url = {https://link.aps.org/doi/10.1103/PRXQuantum.2.020333}
}

@article{ChabaudStellar2020,
  title = {Stellar Representation of Non-Gaussian Quantum States},
  author = {Chabaud, Ulysse and Markham, Damian and Grosshans, Fr\'ed\'eric},
  journal = {Phys. Rev. Lett.},
  volume = {124},
  issue = {6},
  pages = {063605},
  numpages = {6},
  year = {2020},
  month = {Feb},
  publisher = {American Physical Society},
  doi = {10.1103/PhysRevLett.124.063605},
  url = {https://link.aps.org/doi/10.1103/PhysRevLett.124.063605}
}

@article{qutip5,
  title = {QuTiP 5: The Quantum Toolbox in {Python}},
  author = {
    Lambert, Neill and Gigu{`e}re, Eric and Menczel, Paul and Li, Boxi and
    Hopf, Patrick and Su{'a}rez, Gerardo and Gali, Marc and Lishman, Jake and
    Gadhvi, Rushiraj and Agarwal, Rochisha and Galicia, Asier and Shammah, Nathan and
    Nation, Paul and Johansson, J. R. and Ahmed, Shahnawaz and Cross, Simon and
    Pitchford, Alexander and Nori, Franco
  },
  journal = {Physics Reports},
  volume = {1153},
  pages = {1-62},
  year = {2026},
  issn = {0370-1573},
  doi = {10.1016/j.physrep.2025.10.001},
  url = {https://www.sciencedirect.com/science/article/pii/S0370157325002704},
}

@article{johansson2012qutip,
      title = {QuTiP: An open-source Python framework for the dynamics of open quantum systems},
      journal = {Computer Physics Communications},
      volume = {183},
      number = {8},
      pages = {1760-1772},
      year = {2012},
      issn = {0010-4655},
      doi = {https://doi.org/10.1016/j.cpc.2012.02.021},
      url = {https://www.sciencedirect.com/science/article/pii/S0010465512000835},
      author = {J.R. Johansson and P.D. Nation and Franco Nori}
}

@article{Kenfack_2004,
    doi = {10.1088/1464-4266/6/10/003},
    url = {https://doi.org/10.1088/1464-4266/6/10/003},
    year = {2004},
    month = {aug},
    publisher = {},
    volume = {6},
    number = {10},
    pages = {396},
    author = {Anatole Kenfack and Karol Życzkowski},
    title = {Negativity of the Wigner function as an indicator of non-classicality},
    journal = {Journal of Optics B: Quantum and Semiclassical Optics}
    }

@article{JozsaFidelity,
    author = {Richard Jozsa},
    title = {Fidelity for Mixed Quantum States},
    journal = {Journal of Modern Optics},
    volume = {41},
    number = {12},
    pages = {2315--2323},
    year = {1994},
    publisher = {Taylor \& Francis},
    doi = {10.1080/09500349414552171},
}

@article{LachmanCoherences,
  title = {Hierarchies of quantum non-Gaussian coherences for bosonic systems: A theoretical study},
  author = {Lachman, Luk\'a\v{s} and Asenbeck, Beate E. and Boyer, Ambroise and Giri, Priyanka and Urvoy, Alban and Laurat, Julien and Filip, Radim},
  journal = {Phys. Rev. A},
  volume = {111},
  issue = {6},
  pages = {063704},
  numpages = {10},
  year = {2025},
  month = {Jun},
  publisher = {American Physical Society},
  doi = {10.1103/t5c6-byhh},
  url = {https://link.aps.org/doi/10.1103/t5c6-byhh}
}

@article{candes2008enhancing,
  title={Enhancing sparsity by reweighted ℓ 1 minimization},
  author={Candes, Emmanuel J and Wakin, Michael B and Boyd, Stephen P},
  journal={Journal of Fourier analysis and applications},
  volume={14},
  number={5},
  pages={877--905},
  year={2008},
  publisher={Springer}
}

@misc{huet2026,
      title={Industry-ready spin-photon interfaces for hybrid photonic quantum computing}, 
      author={Hêlio Huet and Hubert Lam and Thibaut Pollet and Petr Steindl and Alice Bernard and Albert Adiyatullin and Petr Stepanov and William Hease and Victor Guilloux and Nico Margaria and Joris Verstraten and Raksha Singla and Samuel T. Mister and Anton Pishchagin and Lara Couronné and Samuel Huber and David Sebastian and Duc Duy Tran and Thi Hao Nhi Nguyen and Thi Phuong Do and Joseph Sulpizio and Yann Portella and Kiarn T. Laverick and Thinhinane Bennour and Tomas Alexandre De Sousa and Davide Stefani and Mathias Pont and Maxime Descampeaux and Bianca Scaparra and Martin A. Jacobsen and Klaus D. Jöns and Rinaldo Trotta and Aristide Lemaître and Martina Morassi and Olivier Krebs and Loïc Lanco and Niels Gregersen and Alexia Auffèves and Maria Maffei and Shane Mansfield and Jean Senellart and Thomas Volz and Viviana Villafañe and Stephen C. Wein and Dario A. Fioretto and Sebastien Boissier and Thi Huong Au and Pascale Senellart},
      year={2026},
      eprint={2606.27787},
      archivePrefix={arXiv},
      primaryClass={quant-ph},
      url={https://arxiv.org/abs/2606.27787}, 
}

@article{Daiss2019,
  title = {Single-Photon Distillation via a Photonic Parity Measurement Using Cavity QED},
  author = {Daiss, Severin and Welte, Stephan and Hacker, Bastian and Li, Lin and Rempe, Gerhard},
  journal = {Phys. Rev. Lett.},
  volume = {122},
  issue = {13},
  pages = {133603},
  numpages = {6},
  year = {2019},
  month = {Apr},
  publisher = {American Physical Society},
  doi = {10.1103/PhysRevLett.122.133603},
  url = {https://link.aps.org/doi/10.1103/PhysRevLett.122.133603}
}

@article{Zapletal:21,
author = {Petr Zapletal and Tom Darras and Hanna Le Jeannic and Adrien Cavaill\`{e}s and Giovanni Guccione and Julien Laurat and Radim Filip},
journal = {Optica},
number = {5},
pages = {743--748},
publisher = {Optica Publishing Group},
title = {Experimental Fock-state bunching capability of non-ideal single-photon states},
volume = {8},
month = {May},
year = {2021},
url = {https://opg.optica.org/optica/abstract.cfm?URI=optica-8-5-743},
doi = {10.1364/OPTICA.419230},
}

@article{Zapletal_Filip_2017, title={Multi-copy quantifiers for single-Photon States}, volume={7}, DOI={10.1038/s41598-017-01333-y}, number={1}, journal={Scientific Reports}, author={Zapletal, Petr and Filip, Radim}, year={2017}, month={May}}

@article{Huang2015,
  title   = {Optical Synthesis of Large-Amplitude Squeezed Coherent-State Superpositions with Minimal Resources},
  author  = {Huang, Kun and Le Jeannic, H. and Ruaudel, J. and Verma, V. B. and Shaw, M. D. and Marsili, F. and Nam, S. W. and Wu, E. and Zeng, H. and Jeong, Y.-C. and Filip, R. and Morin, O. and Laurat, J.},
  journal = {Physical Review Letters},
  volume  = {115},
  pages   = {023602},
  year    = {2015},
  doi     = {10.1103/PhysRevLett.115.023602}
}

@article{Etesse2015,
  title   = {Experimental Generation of Squeezed Cat States with an Operation Allowing Iterative Growth},
  author  = {Etesse, Jean and Bouillard, Martin and Kanseri, Bhaskar and Tualle-Brouri, Rosa},
  journal = {Physical Review Letters},
  volume  = {114},
  pages   = {193602},
  year    = {2015},
  doi     = {10.1103/PhysRevLett.114.193602}
}

@article{Sychev2017,
  title   = {Enlargement of Optical Schr{\"o}dinger's Cat States},
  author  = {Sychev, Demid V. and Ulanov, Alexander E. and Pushkina, Anastasia A. and Richards, Matthew W. and Fedorov, Ilya A. and Lvovsky, Alexander I.},
  journal = {Nature Photonics},
  volume  = {11},
  pages   = {379--382},
  year    = {2017},
  doi     = {10.1038/nphoton.2017.57}
}

@article{Christensen2015,
  author = {Christensen, Jesper B. and Reddy, Dileep V. and McKinstrie, C. J. and Rottwitt, Karsten and Raymer, M. G.},
  title = {Temporal mode sorting using dual-stage quantum frequency conversion by asymmetric Bragg scattering},
  journal = {Optics Express},
  volume = {23},
  number = {18},
  pages = {23287--23301},
  year = {2015},
  doi = {10.1364/OE.23.023287}
}

@article{Endo2026,
  title   = {Four-Photon Subtraction from Picosecond Squeezed Vacuum Demonstrating Wigner Negativity without Loss Correction},
  author  = {Endo, Mamoru and Nomura, Takefumi and Sonoyama, Tatsuki and Takahashi, Kazuma and Takasu, Sachiko and Fukuda, Daiji and Kashiwazaki, Takahiro and Inoue, Asuka and Umeki, Takeshi and Nehra, Rajveer and Marek, Petr and Filip, Radim and Takase, Kan and Asavanant, Warit and Furusawa, Akira},
  journal = {Physical Review A},
  volume  = {114},
  pages   = {022618},
  year    = {2026},
  doi     = {10.1103/6tlg-7qgm}
}

@article{TejaQNG2025,
  title = {Quantum non-Gaussian high Fock states of light pulses and their superpositions},
  author = {Teja, G. P. and Kumar, Chandan and Lachman, Luk\'a\ifmmode \check{s}\else \v{s}\fi{} and Filip, Radim},
  journal = {Phys. Rev. Res.},
  volume = {7},
  issue = {3},
  pages = {033272},
  numpages = {17},
  year = {2025},
  month = {Sep},
  publisher = {American Physical Society},
  doi = {10.1103/qyfz-zw2x},
  url = {https://link.aps.org/doi/10.1103/qyfz-zw2x}
}

\end{document}